\documentclass[aps,preprint]{revtex4}%
\usepackage{amsfonts}
\usepackage{amsmath}
\usepackage{amssymb}
\usepackage{graphicx}%
\begin{document}
\preprint{ }
\title{Superfluidity and excitations at unitarity}
\author{Dean Lee}
\affiliation{Department of Physics, North Carolina State University, Raleigh, NC 27695}

\begin{abstract}
We present lattice results for spin-1/2 fermions at unitarity, where the
effective range of the interaction is zero and the scattering length is
infinite. \ We measure the spatial coherence of difermion pairs for a system
of 6, 10, 14, 18, 22, 26 particles with equal numbers of up and down spins in
a periodic cube. \ Using Euclidean time projection, we analyze ground state
properties and transient behavior due to low-energy excitations. \ At
asymptotically large values of $t$ we see long-range order consistent with
spontaneously broken $U(1)$ fermion-number symmetry and a superfluid ground
state. \ At intermediate times we see exponential decay in the $t$-dependent
signal due to an unknown low-energy excitation. \ We probe this low-energy
excitation further by calculating two-particle correlation functions. \ We
find that the excitation has the properties of a chain of particles extending
across the periodic lattice.

\end{abstract}
\maketitle
\tableofcontents

\section{Introduction}

The unitary limit describes a many-body system of nonrelativistic spin-1/2
fermions with zero-range attraction and infinite scattering length. \ While
the unitary limit has a well-defined continuum limit and strong interactions,
at zero temperature it has no intrinsic physical scale other than the
interparticle spacing. \ This implies for example at zero temperature the
energy per particle and pairing gap are both given by the Fermi energy times a
dimensionless constant.

The universal nature of the unitary limit endows it relevance to several areas
of physics and the subject has received much recent interest. \ The ground
state is believed to be superfluid with properties somewhere between a
Bardeen-Cooper-Schrieffer (BCS) fermionic superfluid at weak coupling and a
Bose-Einstein condensate (BEC) of bound dimers at strong coupling
\cite{Eagles:1969PR,Leggett:1980pro,Nozieres:1985JLTP}. \ In solid state
physics it has been suggested that the crossover from BCS fermionic superfluid
to BEC bosonic superfluid describes the pseudogap phase in high-temperature
cuprate superconductors \cite{Chen:2005PhyRep}. \ In atomic physics BCS-BEC
crossover has been studied extensively with trapped ultracold $^{6}$Li and
$^{40}$K atoms. \ The atoms are sufficiently far apart that the effective
range of the interaction is negligible while the scattering length can be
adjusted using a magnetic-field Feshbach resonance
\cite{Tiesinga:1993PRA,Stwalley:1976PRL,Courteille:1998PRL,Inouye:1998Nat}.
\ In nuclear physics the unitary limit is relevant to the properties of cold
dilute neutron matter. \ The neutron scattering length is about $-18$ fm while
the effective range is $2.8$ fm. \ Therefore the unitary limit is
approximately realized when the interparticle spacing is about $10$ fm,
roughly $0.5\%$ of normal nuclear matter density. \ Superfluid neutrons at
around this density may be present in the inner crust of neutron stars
\cite{Pethick:1995di,Lattimer:2004pg}.

In this study we measure the low-energy states of unpolarized spin-1/2
fermions in the unitary limit. \ We use the same lattice projection technique
used in \cite{Lee:2005fk} to measure the ground state energy. \ We start with
a quantum state with the desired quantum numbers and use the Euclidean time
projection operator $e^{-Ht}$ to filter out high-energy excitations. \ We work
with finite systems where the energy spectrum is discrete. \ Therefore we
don't have gapless modes which appear only in the thermodynamic limit. \ After
this filtering we measure spatial correlations\ of the superfluid order
parameter as a function of the projection time $t$. \ At asymptotically large
values of $t$ we see long-range order consistent with spontaneously broken
$U(1)$ fermion-number symmetry and a superfluid ground state. \ At
intermediate times we see exponential decay in the $t$-dependent signal due to
an unknown low-energy excitation. \ We probe this low-energy excitation
further by calculating two-particle correlation functions. \ We find that the
excitation is consistent with a quasi-1D subsystem of particles extending
across the periodic lattice.

\section{Off-diagonal long-range order}

Written in continuum notation the Hamiltonian for the unitary limit is%
\begin{equation}
H=-\frac{1}{2m}\sum_{i=\uparrow,\downarrow}\int d^{3}\vec{r}\;a_{i}^{\dagger
}(\vec{r})\vec{\nabla}^{2}a_{i}(\vec{r})+C\int d^{3}\vec{r}\;a_{\downarrow
}^{\dagger}(\vec{r})a_{\uparrow}^{\dagger}(\vec{r})a_{\uparrow}(\vec
{r})a_{\downarrow}(\vec{r}). \label{hamiltonian}%
\end{equation}
$a_{i}$ and $a_{i}^{\dagger}$ are annihilation and creation operators for
fermions with spin $i$. \ The mass of the fermion is $m$, and the coefficient
$C$ is cutoff dependent. \ We discuss later how in lattice regularization $C$
is tuned to make the s-wave scattering length infinite.

The unitary limit Hamiltonian has a global $U(1)$ fermion-number symmetry%
\begin{equation}
\left[
\genfrac{}{}{0pt}{}{a_{\uparrow}(\vec{r})}{a_{\downarrow}(\vec{r})}%
\right]  \rightarrow e^{i\phi}\left[
\genfrac{}{}{0pt}{}{a_{\uparrow}(\vec{r})}{a_{\downarrow}(\vec{r})}%
\right]  ,
\end{equation}
where $\phi$ is any real constant. \ It also has a global $SU(2)$ spin
symmetry%
\begin{equation}
\left[
\genfrac{}{}{0pt}{}{a_{\uparrow}(\vec{r})}{a_{\downarrow}(\vec{r})}%
\right]  \rightarrow e^{i\vec{\phi}\cdot\vec{\sigma}}\left[
\genfrac{}{}{0pt}{}{a_{\uparrow}(\vec{r})}{a_{\downarrow}(\vec{r})}%
\right]  ,
\end{equation}
where $\vec{\sigma}$ denotes the $2\times2$ Pauli spin matrices and $\vec
{\phi}$ is any real constant three-component vector. \ Since there is no
coupling between intrinsic spin and orbital angular momentum, this $SU(2)$
symmetry should be regarded as an internal symmetry decoupled from spatial rotations.

The lowest-dimensional local bosonic operator that can be constructed from the
annihilation field operators is%
\begin{equation}
\psi^{2}(\vec{r})=a_{\uparrow}(\vec{r})a_{\downarrow}(\vec{r}).
\end{equation}
We note that $\psi^{2}$ is invariant under the $SU(2)$ spin symmetry but phase
rotates under the $U(1)$ fermion-number symmetry,
\begin{equation}
\psi^{2}(\vec{r})\rightarrow e^{2i\phi}\psi^{2}(\vec{r}).
\end{equation}
Therefore if there is some critical temperature below which $\psi^{2}$ has
long-range spatial correlations,
\begin{equation}
\lim_{\left\vert \vec{r}\right\vert \rightarrow\infty}\left\langle
\psi^{2\dagger}(\vec{r})\psi^{2}(\vec{0})\right\rangle \neq0,
\end{equation}
then the $U(1)$ fermion-number symmetry is spontaneously broken. \ While there
are many different ways to characterize superfluid behavior, this condition of
off-diagonal long-range order \cite{Penrose:1956,Gorkov:1958} is usually
regarded as the standard definition for superfluidity.

\section{Measured observables in continuum notation}

In order to highlight the physics content of the lattice calculation we
summarize the measured observables in continuum notation. \ We refer to a
state with $N_{\uparrow}$ up-spin fermions and $N_{\downarrow}$ down-spin
fermions as an $N_{\uparrow},N_{\downarrow}$ state. \ We also specify the
total momentum $\vec{P}$ and total spin $S$ of the $SU(2)$ spin
representation. \ Let $\left\vert \Psi_{0}^{\text{free}}\right\rangle $ be the
free Fermi ground state for the $N,N$ system with $\vec{P}=\vec{0}$ and $S=0$.
\ We filter out high-energy states using the Euclidean-time projection
operator $e^{-Ht}$,
\begin{equation}
\left\vert \Psi(t)\right\rangle =e^{-Ht}\left\vert \Psi_{0}^{\text{free}%
}\right\rangle .
\end{equation}
With the projected states $\left\vert \Psi(t)\right\rangle $ we measure the
correlation function $G_{\psi^{2}}(\vec{r},t_{1},t_{2})$,
\begin{equation}
G_{\psi^{2}}(\vec{r},t_{1},t_{2})=\frac{1}{\Gamma}\frac{\left\langle
\Psi(t_{1})\right\vert \psi^{2\dagger}(\vec{r})\psi^{2}(\vec{0})\left\vert
\Psi(t_{2})\right\rangle }{\left\langle \Psi(t_{1})\right.  \left\vert
\Psi(t_{2})\right\rangle }, \label{difermioncorr}%
\end{equation}
where $\Gamma$ is a renormalization coefficient set by the condition%
\begin{equation}
\lim_{t_{1},t_{2}\rightarrow\infty}G_{\psi^{2}}(\vec{0},t_{1},t_{2})=1.
\end{equation}

We consider only finite systems where the energy spectrum is discrete. \ For
large $t$ we have the asymptotic form%
\begin{equation}
\left\vert \Psi(t)\right\rangle =c_{0}e^{-E_{0}t}\left\vert \Psi
_{0}\right\rangle +c_{1}e^{-E_{1}t}\left\vert \Psi_{1}\right\rangle +\cdots.
\label{asymptotic}%
\end{equation}
$\left\vert \Psi_{0}\right\rangle $ is the normalized ground state with energy
$E_{0}$, and $\left\vert \Psi_{1}\right\rangle $ is the normalized first
excited state with energy $E_{1}$. \ We anticipate the possibility of
degeneracy in the low-energy spectrum corresponding with two-particle
excitations with momenta $\vec{k}$ and $-\vec{k}$. \ Since we use a periodic
cube with lattice regularization, quantum states in the same irreducible
representation of the cubic lattice symmetry group $SO(3,\mathbb{Z})$ are
exactly degenerate in energy. \ But there might also be approximately
degenerate excited states whose energy differences are not adequately resolved
by the finite values of $t$ measured. \ We use the notation $\left\vert
\Psi_{1,i}\right\rangle $ for the complete set of normalized degenerate
excited states with energy $E_{1}$. $\ $Using the asymptotic form
(\ref{asymptotic}) we get%
\begin{align}
G_{\psi^{2}}(\vec{r},t_{1},t_{2})  &  =A_{00}(\vec{r})+A_{01}(\vec
{r})e^{-(E_{1}-E_{0})t_{2}}+A_{10}(\vec{r})e^{-(E_{1}-E_{0})t_{1}}\nonumber\\
&  +A_{11}(\vec{r})e^{-(E_{1}-E_{0})(t_{1}+t_{2})}+\cdots\label{G_asymp}%
\end{align}
where%
\begin{equation}
A_{00}(\vec{r})=\frac{1}{\Gamma}\left\langle \Psi_{0}\right\vert
\psi^{2\dagger}(\vec{r})\psi^{2}(\vec{0})\left\vert \Psi_{0}\right\rangle ,
\label{A00}%
\end{equation}

\begin{equation}
A_{01}(\vec{r})=\frac{1}{\Gamma}\sum_{i}\frac{c_{0}^{\ast}c_{1,i}}{\left\vert
c_{0}\right\vert ^{2}}\left\langle \Psi_{0}\right\vert \psi^{2\dagger}(\vec
{r})\psi^{2}(\vec{0})\left\vert \Psi_{1,i}\right\rangle , \label{A01}%
\end{equation}%
\begin{equation}
A_{10}(\vec{r})=\frac{1}{\Gamma}\sum_{i}\frac{c_{1,i}^{\ast}c_{0}}{\left\vert
c_{0}\right\vert ^{2}}\left\langle \Psi_{1,i}\right\vert \psi^{2\dagger}%
(\vec{r})\psi^{2}(\vec{0})\left\vert \Psi_{0}\right\rangle , \label{A10}%
\end{equation}%
\begin{equation}
A_{11}(\vec{r})=\frac{1}{\Gamma}\sum_{i,i^{\prime}}\frac{c_{1,i}^{\ast
}c_{1,i^{\prime}}}{\left\vert c_{0}\right\vert ^{2}}\left\langle \Psi
_{1,i}\right\vert \psi^{2\dagger}(\vec{r})\psi^{2}(\vec{0})\left\vert
\Psi_{1,i^{\prime}}\right\rangle -\frac{1}{\Gamma}\sum_{i}\frac{\left\vert
c_{1,i}\right\vert ^{2}}{\left\vert c_{0}\right\vert ^{2}}\left\langle
\Psi_{0}\right\vert \psi^{2\dagger}(\vec{r})\psi^{2}(\vec{0})\left\vert
\Psi_{0}\right\rangle . \label{A11}%
\end{equation}
From the definition of $\Gamma$ it follows that $A_{00}(\vec{0})=1$. \ In
order to determine whether the strongest transient signal comes from
$A_{01}(\vec{r})$ and $A_{10}(\vec{r})$ or $A_{11}(\vec{r})$ we will calculate
$G_{\psi^{2}}(\vec{r},t_{1},t_{2})$ for two different large time combinations.
\ In one case we let $t_{1}=t_{2}$ and send both to infinity. \ In the other
case we let $t_{2}$ be large but fixed and take the limit as $t_{1}$ goes to
infinity. \ By analyzing the time dependence we extract the energy difference
$E_{1}-E_{0}$, and from the spatial dependence we measure the momentum of the
particles in $\left\vert \Psi_{1,i}\right\rangle $ which couple to
$\psi^{2\dagger}(\vec{r})\psi^{2}(\vec{0})$.

\section{Lattice formalism}

Throughout our discussion of the lattice calculation we use dimensionless
parameters and operators which correspond with physical values multiplied by
the appropriate power of the spatial lattice spacing $a$. \ When we need to
specify quantities in physical units we write the subscript `phys'. \ In our
notation the four-component integer vector $\vec{n}$ labels the lattice sites
of a $3+1$ dimensional lattice with dimensions $L^{3}\times L_{t}$. $\ \vec
{n}_{s}$ gives the spatial part of $\vec{n}$, while $n_{t}$ is the time
component. We write$\ \hat{l}_{s}=\hat{1}$, $\hat{2}$, $\hat{3}$ for the
spatial lattice unit vectors and $\hat{0}$ for the temporal lattice unit
vector. \ The temporal lattice spacing is given by $a_{t}$, and $\alpha
_{t}=a_{t}/a$ is the ratio of the temporal to spatial lattice spacing. \ We
also define $h=\alpha_{t}/(2m)$, where $m$ is the fermion mass in lattice units.

We briefly discuss four different lattice formulations: \ the transfer matrix
formalism with and without auxiliary fields, and the path integral formalism
with and without auxiliary fields. While in the main calculation we use only
the transfer matrix formalism with auxiliary fields, the other formulations
are useful to provide numerical checks of the simulation data. \ The four
formulations agree exactly even for nonzero spatial and temporal lattice spacings.

The first formulation we consider is the path integral formalism with
auxiliary fields. \ This has been used in varying forms in several grand
canonical mean field calculations and lattice simulations at nonzero
temperature
\cite{Chen:2003vy,Lee:2004qd,Lee:2004si,Wingate:2005xy,Bulgac:2005a,Lee:2005is,Lee:2005it}%
. \ We let $c_{i}(\vec{n})$ and $c_{i}^{\ast}(\vec{n})$ be anticommuting
Grassmann fields for spin $i$, and $s(\vec{n})$ be an auxiliary
Hubbard-Stratonovich field. \ Let $\mathcal{Z}$ be the partition function%

\begin{equation}
\mathcal{Z}\propto\int DsDcDc^{\ast}\exp\left[  -S\left(  s,c,c^{\ast}\right)
\right]  ,
\end{equation}
where%
\begin{equation}
Ds=\prod_{\vec{n}}ds(\vec{n}),
\end{equation}%
\begin{equation}
DcDc^{\ast}=\prod_{\vec{n},i}dc_{i}(\vec{n})dc_{i}^{\ast}(\vec{n}).
\end{equation}
We take as our standard lattice path integral action%
\begin{align}
S(s,c,c^{\ast})=  &  \sum_{\vec{n},i}\left[  c_{i}^{\ast}(\vec{n})c_{i}%
(\vec{n}+\hat{0})-e^{\sqrt{-C\alpha_{t}}s(\vec{n})+\frac{C\alpha_{t}}{2}%
}(1-6h)c_{i}^{\ast}(\vec{n})c_{i}(\vec{n})\right] \nonumber\\
&  -h\sum_{\vec{n},l_{s},i}\left[  c_{i}^{\ast}(\vec{n})c_{i}(\vec{n}+\hat
{l}_{s})+c_{i}^{\ast}(\vec{n})c_{i}(\vec{n}-\hat{l}_{s})\right]  +\frac{1}%
{2}\sum_{\vec{n}}\left[  s(\vec{n})\right]  ^{2}.
\end{align}
If we include a chemical potential then the action becomes%
\begin{align}
S(s,c,c^{\ast})=  &  \sum_{\vec{n},i}\left[  c_{i}^{\ast}(\vec{n})c_{i}%
(\vec{n}+\hat{0})-e^{\sqrt{-C\alpha_{t}}s(\vec{n})+\frac{C\alpha_{t}}{2}%
}e^{\mu\alpha_{t}}(1-6h)c_{i}^{\ast}(\vec{n})c_{i}(\vec{n})\right] \nonumber\\
&  -he^{\mu\alpha_{t}}\sum_{\vec{n},l_{s},i}\left[  c_{i}^{\ast}(\vec{n}%
)c_{i}(\vec{n}+\hat{l}_{s})+c_{i}^{\ast}(\vec{n})c_{i}(\vec{n}-\hat{l}%
_{s})\right]  +\frac{1}{2}\sum_{\vec{n}}\left[  s(\vec{n})\right]  ^{2}.
\end{align}
From this point on it is easy to follow how the chemical potential enters into
the other formulations. \ Therefore we simplify the discussion by setting the
chemical potential to zero.

In order to connect the path integral with the transfer matrix formalism, we
use the correspondence \cite{Creutz:1988wv,Creutz:1999zy}%
\begin{align}
&  Tr\left\{  \colon F_{L_{t}-1}\left[  a_{i^{\prime}}^{\dagger}(\vec{n}%
_{s}^{\prime}),a_{i}(\vec{n}_{s})\right]  \colon\times\cdots\times\colon
F_{0}\left[  a_{i^{\prime}}^{\dagger}(\vec{n}_{s}^{\prime}),a_{i}(\vec{n}%
_{s})\right]  \colon\right\} \nonumber\\
&  =\int DcDc^{\ast}\exp\left\{  \sum_{n_{t}=0}^{L_{t}-1}\sum_{\vec{n}_{s}%
,i}c_{i}^{\ast}(\vec{n}_{s},n_{t})\left[  c_{i}(\vec{n}_{s},n_{t})-c_{i}%
(\vec{n}_{s},n_{t}+1)\right]  \right\} \nonumber\\
&  \qquad\qquad\qquad\times\prod_{n_{t}=0}^{L_{t}-1}F_{n_{t}}\left[
c_{i^{\prime}}^{\ast}(\vec{n}_{s}^{\prime},n_{t}),c_{i}(\vec{n}_{s}%
,n_{t})\right]  \label{correspondence}%
\end{align}
where $c_{i}(\vec{n}_{s},L_{t})=-c_{i}(\vec{n}_{s},0).$ \ We use $a_{i}%
(\vec{n}_{s})$ and $a_{i}^{\dagger}(\vec{n}_{s})$ to denote the fermion
annihilation and creation operators respectively for spin $i$. \ The $\colon$
symbols in the top line of (\ref{correspondence}) indicate normal ordering.
Let us define the transfer matrix operator
\begin{equation}
M_{n_{t}}(s)\equiv\colon\exp\left\{
\begin{array}
[c]{c}%
\sum_{\vec{n}_{s},i}\left[  e^{\sqrt{-C\alpha_{t}}s(\vec{n}_{s},n_{t}%
)+\frac{C\alpha_{t}}{2}}(1-6h)-1\right]  a_{i}^{\dagger}(\vec{n}_{s}%
)a_{i}(\vec{n}_{s})\\
+h\sum_{\vec{n}_{s},l_{s},i}\left[  a_{i}^{\dagger}(\vec{n}_{s})a_{i}(\vec
{n}_{s}+\hat{l}_{s})+a_{i}^{\dagger}(\vec{n}_{s})a_{i}(\vec{n}_{s}-\hat{l}%
_{s})\right]
\end{array}
\right\}  \colon. \label{auxiliary_transfer}%
\end{equation}
We can write the partition function as
\begin{equation}
\mathcal{Z}\propto%
{\displaystyle\int}
Ds\;Tr\left\{  M_{L_{t}-1}(s)\times\cdots\times M_{0}(s)\right\}  \exp\left\{
\frac{1}{2}\sum_{\vec{n}}\left[  s(\vec{n})\right]  ^{2}\right\}  .
\end{equation}

\begin{figure}
[ptb]
\begin{center}
\includegraphics[
height=1.2107in,
width=5.028in
]%
{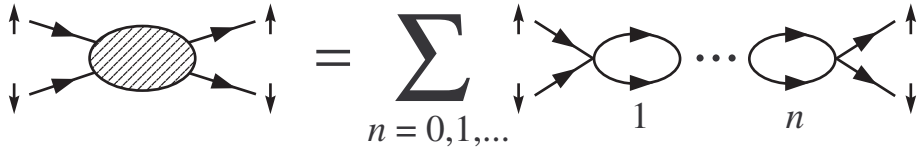}%
\caption{Two-particle scattering bubble chain diagrams}%
\label{twotwo}%
\end{center}
\end{figure}

We now remove the auxiliary field. \ Integrating over $s(\vec{n})$ in the path
integral we obtain \cite{Borasoy:2005yc}%
\begin{equation}
\mathcal{Z}\propto\int DcDc^{\ast}\exp\left[  -S\left(  c,c^{\ast}\right)
\right]  ,
\end{equation}%
\begin{align}
S(c,c^{\ast})=  &  \sum_{\vec{n},i}\left[  c_{i}^{\ast}(\vec{n})c_{i}(\vec
{n}+\hat{0})-(1-6h)c_{i}^{\ast}(\vec{n})c_{i}(\vec{n})\right] \nonumber\\
&  -h\sum_{\vec{n},l_{s},i}\left[  c_{i}^{\ast}(\vec{n})c_{i}(\vec{n}+\hat
{l}_{s})+c_{i}^{\ast}(\vec{n})c_{i}(\vec{n}-\hat{l}_{s})\right] \nonumber\\
&  -\left(  e^{-C\alpha_{t}}-1\right)  (1-6h)^{2}\sum_{\vec{n}}c_{\downarrow
}^{\ast}(\vec{n})c_{\uparrow}^{\ast}(\vec{n})c_{\uparrow}(\vec{n}%
)c_{\downarrow}(\vec{n}). \label{path_nonaux}%
\end{align}
This is probably the simplest formulation for computing Feynman diagrams and
the most convenient for setting the lattice interaction coefficient $C$. \ As
shown in \cite{Lee:2004qd} the procedure for setting $C$ involves summing the
bubble diagrams shown in FIG. \ref{twotwo}, locating the pole in the
scattering amplitude, and comparing with L\"{u}scher's formula for energy
levels in a finite periodic cube \cite{Luscher:1986pf,Beane:2003da},%

\begin{equation}
E_{\text{pole}}=\dfrac{4\pi a_{\text{scatt}}}{mL^{3}}[1-c_{1}\frac
{a_{\text{scatt}}}{L}+c_{2}\frac{a_{\text{scatt}}^{2}}{L^{2}}+\cdots],
\label{lus}%
\end{equation}
where $c_{1}=-2.837297,$ $c_{2}=6.375183$. \ Taking the limit $a_{\text{scatt}%
}\rightarrow\infty$ we get the interaction coefficient for the unitary limit.%

We can use (\ref{correspondence}) again to derive the corresponding transfer
matrix formalism without auxiliary fields,%
\begin{equation}
\mathcal{Z}\propto Tr\left\{  M_{L_{t}-1}\times\cdots\times M_{0}\right\}  ,
\end{equation}
where%
\begin{equation}
M_{n_{t}}\equiv\colon\exp\left\{
\begin{array}
[c]{c}%
-6h\sum_{\vec{n}_{s},i}a_{i}^{\dagger}(\vec{n}_{s})a_{i}(\vec{n}_{s}%
)+h\sum_{\vec{n}_{s},l_{s},i}\left[  a_{i}^{\dagger}(\vec{n}_{s})a_{i}(\vec
{n}_{s}+\hat{l}_{s})+a_{i}^{\dagger}(\vec{n}_{s})a_{i}(\vec{n}_{s}-\hat{l}%
_{s})\right] \\
+\left(  e^{-C\alpha_{t}}-1\right)  (1-6h)^{2}\sum_{\vec{n}_{s}}a_{\downarrow
}^{\dagger}(\vec{n}_{s})a_{\uparrow}^{\dagger}(\vec{n}_{s})a_{\uparrow}%
(\vec{n}_{s})a_{\downarrow}(\vec{n}_{s})
\end{array}
\right\}  \colon. \label{noauxiliary_transfer}%
\end{equation}
Since there is no time-dependent auxiliary field, $M_{n_{t}}$ is the same for
each time step $n_{t}$. \ This formulation is useful in few-body systems with
no sign problem. \ In \cite{Borasoy:2005yc} it was used to calculate the
binding energy of the $SU(4)$ Wigner-symmetric triton.

\section{Transfer matrix projection method}

We use the transfer matrix projection method introduced in \cite{Lee:2005fk}.
\ In this paper we consider the values $N=1,3,5,7,9,11,13$. \ For each spin we
fill the momentum states comprising $\left\vert \Psi_{0}^{\text{free}%
}\right\rangle $ in the order shown in Table 1. \ We observe that for each
value of $N$, $\left\vert \Psi_{0}^{\text{free}}\right\rangle $ has zero total
momentum and zero total spin. \ For $N=7$ we have the special case where
$\left\vert \Psi_{0}^{\text{free}}\right\rangle $ is also invariant under
$SO(3,\mathbb{Z})$ rotations.%

\[%
\genfrac{}{}{0pt}{}{%
\begin{tabular}
[c]{|c|c|}\hline
$N$ & additional momenta filled\\\hline
$1$ & $\left\langle 0,0,0\right\rangle $\\\hline
$3$ & $\left\langle \frac{2\pi}{L},0,0\right\rangle ,\left\langle -\frac{2\pi
}{L},0,0\right\rangle $\\\hline
$5$ & $\left\langle 0,\frac{2\pi}{L},0\right\rangle ,\left\langle
0,-\frac{2\pi}{L},0\right\rangle $\\\hline
$7$ & $\left\langle 0,0,\frac{2\pi}{L}\right\rangle ,\left\langle
0,0,-\frac{2\pi}{L}\right\rangle $\\\hline
$9$ & $\left\langle \frac{2\pi}{L},\frac{2\pi}{L},0\right\rangle ,\left\langle
-\frac{2\pi}{L},-\frac{2\pi}{L},0\right\rangle $\\\hline
$11$ & $\left\langle \frac{2\pi}{L},-\frac{2\pi}{L},0\right\rangle
,\left\langle -\frac{2\pi}{L},\frac{2\pi}{L},0\right\rangle $\\\hline
$13$ & $\left\langle 0,\frac{2\pi}{L},\frac{2\pi}{L}\right\rangle
,\left\langle 0,-\frac{2\pi}{L},-\frac{2\pi}{L}\right\rangle $\\\hline
\end{tabular}
}{\text{Table 1. \ Filling sequence of momentum states}}%
\]
\bigskip

Using the auxiliary field transfer matrix defined in (\ref{auxiliary_transfer}%
) we construct%
\begin{equation}
\left\vert \Psi(t),s\right\rangle \equiv M_{n_{t}-1}(s)\times\cdots\times
M_{0}(s)\left\vert \Psi_{0}^{\text{free}}\right\rangle ,
\end{equation}
where $t=n_{t}a_{t}$. \ This is the analog of
\begin{equation}
\left\vert \Psi(t)\right\rangle =e^{-Ht}\left\vert \Psi_{0}^{\text{free}%
}\right\rangle
\end{equation}
defined above in the continuum notation. \ We compute spatial correlations of
the difermion pair operator $\psi^{2}(\vec{n}_{s})=a_{\uparrow}(\vec{n}%
_{s})a_{\downarrow}(\vec{n}_{s})$,%
\begin{equation}
G_{\psi^{2}}^{\text{bare}}(\vec{n}_{s},t_{1},t_{2})=\frac{%
{\displaystyle\int}
Ds\;\left\langle \Psi(t_{1}),s\right\vert \psi^{2\dagger}(\vec{n}_{s})\psi
^{2}(\vec{0})\left\vert \Psi(t_{2}),s\right\rangle \exp\left\{  -\frac{1}%
{2}\sum_{\vec{n}}\left[  s(\vec{n})\right]  ^{2}\right\}  }{%
{\displaystyle\int}
Ds\;\left\langle \Psi(t_{1}),s\right.  \left\vert \Psi(t_{2}),s\right\rangle
\exp\left\{  -\frac{1}{2}\sum_{\vec{n}}\left[  s(\vec{n})\right]
^{2}\right\}  }.
\end{equation}
The inner products in the numerator and denominator are to be defined shortly.
\ We use the superscript `bare' since matrix elements of the composite
operators $\psi^{2}$ and $\psi^{2\dagger}$ diverge in the continuum limit.
\ We take care of this renormalization by rescaling the correlation function,%
\begin{equation}
G_{\psi^{2}}(\vec{n}_{s},t_{1},t_{2})=\frac{1}{\Gamma}G_{\psi^{2}%
}^{\text{bare}}(\vec{n}_{s},t_{1},t_{2}),
\end{equation}
where \
\begin{equation}
\Gamma=\lim_{t_{1},t_{2}\rightarrow\infty}G_{\psi^{2}}^{\text{bare}}(\vec
{0},t_{1},t_{2}).
\end{equation}

$M_{n_{t}}(s)$ consists entirely of single-body operators interacting with the
background auxiliary field and has no direct interactions between particles.
\ This may not be obvious from the complicated form for $M_{n_{t}}(s)$ in
(\ref{auxiliary_transfer}). \ To see this more clearly we pretend for the
moment that the $N$ up-spin particles and $N$ down-spin particles are all
distinguishable. \ We label the newly distinguishable particles with the label
$X=1,2,\cdots,2N-1,2N.$ \ This error in quantum statistics has no effect on
the final answer if the initial and final state wavefunctions are completely
antisymmetric in the up-spin and down-spin variables. \ Since we have exactly
one particle of each type $X$, the normal-ordered operator $M_{n_{t}}(s)$ can
be factorized as a product of terms of the form
\begin{equation}
M_{n_{t}}(s)=%
{\displaystyle\prod\limits_{X}}
M_{n_{t}}^{X}(s),
\end{equation}
where%
\begin{align}
M_{n_{t}}^{X}(s)  &  =1+\left[  e^{\sqrt{-C\alpha_{t}}s(\vec{n}_{s}%
,n_{t})+\frac{C\alpha_{t}}{2}}(1-6h)-1\right]  \sum_{\vec{n}_{s}}%
a_{X}^{\dagger}(\vec{n}_{s})a_{X}(\vec{n}_{s})\nonumber\\
&  +h\sum_{\vec{n}_{s},l_{s}}\left[  a_{X}^{\dagger}(\vec{n}_{s})a_{X}(\vec
{n}_{s}+\hat{l}_{s})+a_{X}^{\dagger}(\vec{n}_{s})a_{X}(\vec{n}_{s}-\hat{l}%
_{s})\right]  . \label{transfermatrix_X}%
\end{align}
If the particle stays at the same spatial lattice site from time step $n_{t}$
to $n_{t}+1$, then the corresponding matrix element of $M_{n_{t}}^{X}(s)$ is%
\begin{equation}
e^{\sqrt{-C\alpha_{t}}s(\vec{n})+\frac{C\alpha_{t}}{2}}(1-6h).
\end{equation}
If the particle hops to a neighboring lattice site from time step $n_{t}$ to
$n_{t}+1$ then the corresponding matrix element of $M_{n_{t}}^{X}(s)$ is $h$.
\ All other elements of $M_{n_{t}}^{X}(s)$ are zero.

We can therefore compute the full $2N$-body matrix element as the square of
the determinant of the single-particle matrix elements,%
\begin{align}
\left\langle \Psi(t_{1}),s\right.  \left\vert \Psi(t_{2}),s\right\rangle  &
\equiv\left\langle \Psi_{0}^{\text{free}}\right\vert M_{L_{t}-1}%
(s)\times\cdots\times M_{0}(s)\left\vert \Psi_{0}^{\text{free}}\right\rangle
\nonumber\\
&  =\left[  \det M(s,t_{1}+t_{2})\right]  ^{2},
\end{align}%
\begin{equation}
M_{ij}(s,t_{1}+t_{2})=\left\langle p_{i}^{X}\right\vert M_{L_{t}-1}%
^{X}(s)\times\cdots\times M_{0}^{X}(s)\left\vert p_{j}^{X}\right\rangle ,
\end{equation}
where $i,j$ go from $1$ to $N$ and $t_{1}+t_{2}=L_{t}a_{t}$. \ The states
$\left\vert p_{j}^{X}\right\rangle $ are the single-particle momentum states
for the up spins (or down spins) comprising our Slater determinant initial and
final state $\left\vert \Psi_{0}^{\text{free}}\right\rangle $. \ The square of
the determinant arises from the fact that we have the same transfer matrix
elements and the same momentum states $\left\vert p_{j}^{X}\right\rangle $ for
both up and down spins. \ Since the square of the determinant is nonnegative,
there is no sign problem.

We sample configurations according to the weight%
\begin{equation}
\exp\left\{  -\frac{1}{2}\sum_{\vec{n}}\left[  s(\vec{n})\right]  ^{2}%
+2\ln\left[  \left\vert \det M(s,t_{1}+t_{2})\right\vert \right]  \right\}  .
\end{equation}
The updating procedure is done using hybrid Monte Carlo \cite{Duane:1987de}.
\ This involves computing molecular dynamics trajectories for%
\begin{equation}
H(s,p)=\frac{1}{2}\sum_{\vec{n}}\left(  p(\vec{n})\right)  ^{2}+V(s),
\end{equation}
where $p(\vec{n})$ is the conjugate momentum for $s(\vec{n})$ and%
\begin{equation}
V(s)=\frac{1}{2}\sum_{\vec{n}}\left(  s(\vec{n})\right)  ^{2}-2\ln\left[
\left\vert \det M(s,t_{1}+t_{2})\right\vert \right]  .
\end{equation}
More details of the updating procedure are given in \cite{Lee:2005fk}. \ For
each configuration we compute the observable
\begin{equation}
O(\vec{n}_{s},t_{1},t_{2},s)=\frac{\;\left\langle \Psi(t_{1}),s\right\vert
\psi^{2\dagger}(\vec{n}_{s})\psi^{2}(\vec{0})\left\vert \Psi(t_{2}%
),s\right\rangle }{\left\langle \Psi(t_{1}),s\right.  \left\vert \Psi
(t_{2}),s\right\rangle }.
\end{equation}
This can be written in terms of the matrix $M(s,t_{1}+t_{2})$ as we now show. \ 

We start with%
\begin{align}
&  \left\langle \Psi(t_{1}),s\right\vert \psi^{2\dagger}(\vec{n}_{s})\psi
^{2}(\vec{0})\left\vert \Psi(t_{2}),s\right\rangle \nonumber\\
&  \equiv\left\langle \Psi_{0}^{\text{free}}\right\vert M_{L_{t}-1}%
(s)\times\cdots\times M_{n_{t_{2}}}(s)\psi^{2\dagger}(\vec{n}_{s})\psi
^{2}(\vec{0})M_{n_{t_{2}}-1}(s)\times\cdots\times M_{0}(s)\left\vert \Psi
_{0}^{\text{free}}\right\rangle
\end{align}
where $t_{1}+t_{2}=L_{t}a_{t}$ and $t_{2}=n_{t_{2}}a_{t}$. \ For each spin the
$\psi^{2\dagger}(\vec{n}_{s})\psi^{2}(\vec{0})$ operator replaces one\ matrix
element from $M(s,t_{1}+t_{2})$, call it the entry in the $k$th row and $l$th
column, with the new matrix element%
\begin{equation}
g_{kl}(\vec{n}_{s},s,t_{1},t_{2})=\left\langle p_{k}^{X}\right\vert
M_{L_{t}-1}^{X}(s)\times\cdots\times M_{n_{t_{2}}}^{X}(s)a_{X}^{\dagger}%
(\vec{n}_{s})a_{X}(\vec{0})M_{n_{t_{2}}-1}^{X}(s)\times\cdots\times M_{0}%
^{X}(s)\left\vert p_{l}^{X}\right\rangle .
\end{equation}
Instead of the full $N\times N$ matrix determinant $\det M(s,t_{1}+t_{2})$, in
this case we get $g_{kl}(\vec{n}_{s},s,t_{1},t_{2})$ times the entry in the
$k$th row and $l$th column of the cofactor matrix of $M(s,t_{1}+t_{2})$.
\ This cofactor matrix element is $(-1)^{k+l}$ times the determinant of the
remaining ($N-1)\times(N-1)$ matrix with the $k$th row and $l$th column of
$M(s,t_{1}+t_{2})$ deleted, and it can be calculated as
\begin{equation}
\det M(s,t_{1}+t_{2})\times\left[  M^{-1}(s,t_{1}+t_{2})\right]  _{lk},
\end{equation}
where $M^{-1}$ is the matrix inverse of $M$. \ Summing over $k$ and $l$ and
squaring the result for the two spins, we get
\begin{align}
&  \left\langle \Psi(t_{1}),s\right\vert \psi^{2\dagger}(\vec{n}_{s})\psi
^{2}(\vec{0})\left\vert \Psi(t_{2}),s\right\rangle \nonumber\\
&  =\left\{  \det M(s,t_{1}+t_{2})\times\sum_{k,l}g_{kl}(\vec{n}_{s}%
,s,t_{1},t_{2})\left[  M^{-1}(s,t_{1}+t_{2})\right]  _{lk}\right\}  ^{2}.
\end{align}
Therefore our observable is
\begin{equation}
O(\vec{n}_{s},t_{1},t_{2},s)=\left\{  \sum_{k,l}g_{kl}(\vec{n}_{s}%
,s,t_{1},t_{2})\left[  M^{-1}(s,t_{1}+t_{2})\right]  _{lk}\right\}  ^{2}.
\end{equation}
By measuring the ensemble average of $O(\vec{n}_{s},t_{1},t_{2},s)$ we get an
unbiased estimate of $G_{\psi^{2}}^{\text{bare}}(\vec{n}_{s},t_{1},t_{2}).$

\section{Lattice parameters and error estimates}

Our choice for the physical values of the fermion mass and lattice spacings
are irrelevant to the universal physics of the unitary limit. \ Nevertheless
we must assign values to these parameters, and the values we choose are
motived by the dilute neutron system. We use a fermion mass of $939$ MeV and
lattice spacings $a=(50$ MeV$)^{-1}$, $a_{t}=(24$ MeV$)^{-1}$. \ For these
parameters we find in the unitary limit, $C_{\text{phys}}=-1.203\times10^{-4}$
MeV$^{-2}$.

For each simulation we compute roughly $2\times10^{5}$ hybrid Monte Carlo
trajectories, split across four processors running completely independent
trajectories. \ Averages and errors are computed by comparing the results of
each processor. \ We use double precision arithmetic to compute $\det
M(s,t_{1}+t_{2})$ and $O(\vec{n}_{s},t_{1},t_{2},s)$. \ All systematic errors
produced by double precision roundoff error and exceptional configurations are
monitored in the following way. \ We introduce a small positive parameter
$\epsilon$ and reject any hybrid Monte Carlo trajectories which generate a
configuration with%
\begin{equation}
\left\vert \det M(s,t_{1}+t_{2})\right\vert <\epsilon^{N}\prod
\limits_{i=1,...,N}\left\vert M_{ii}(s,t_{1}+t_{2})\right\vert \text{.}%
\end{equation}
We then take the limit $\epsilon\rightarrow0^{+}$ to determine if
poorly-conditioned matrices make any detectable contribution to our
observables. \ We consider values for $\epsilon$ as small as $10^{-7}$. \ If
as we take $\epsilon\rightarrow0^{+}$ any systematic error can be detected
above the stochastic error level, then we throw out the measurement and do not
include it in the final results. \ The error bars we present are therefore
estimates of the total error for each lattice system. \ There are no
additional errors other than the lattice spacing dependence.

In the unitary limit our Euclidean variables $t_{1}$ and $t_{2}$ can be
replaced by the dimensionless combinations $\frac{t_{1}}{mL^{2}}$ and
$\frac{t_{2}}{mL^{2}}$. \ More convenient though is to use the dimensionless
combinations $E_{F}t_{1}$ and $E_{F}t_{2}$, where $E_{F}$ is the Fermi energy%
\begin{equation}
E_{F}=\frac{k_{F}^{2}}{2m}=\frac{\left(  6\pi^{2}N\right)  ^{2/3}}{2mL^{2}%
}\approx7.596\frac{N^{2/3}}{mL^{2}}.
\end{equation}
At unitarity we can reach the continuum limit at fixed particle number by
increasing $L$, the length of the periodic cube in lattice units. \ This may
seem an unusual way to take the continuum limit. \ It works only in scale
invariant theories such as the unitary limit or noninteracting fermions where
the only physical scale is the interparticle spacing. \ Since the number of
particles is fixed, the spacing between particles as measured in lattice units
increases as we increase $L$. \ Therefore $L\rightarrow\infty$ corresponds
with the continuum limit. \ By comparing results for different $L$ we obtain
an error estimate for the extrapolation to the continuum limit.

\section{Numerical crosschecks}

To test the lattice codes, we have run simulations for several systems where
the final answer can be calculated accurately by alternative means. \ For the
first test we consider the noninteracting fermion system for the $7,7$ system
with total momentum $\vec{P}=\vec{0}$ and total spin $S=0$. \ We take $L=4$
and set $L_{t}$ large enough to extract the limit \
\begin{equation}
\lim_{t_{1},t_{2}\rightarrow\infty}G_{\psi^{2}}^{\text{bare}}(\vec{n}%
_{s},t_{1},t_{2})
\end{equation}
for $\vec{n}_{s}=\left\langle n_{x},0,0\right\rangle $ with $n_{x}=0,1,2$.
\ We perform the numerical check using the path integral action $S(c,c^{\ast
})$ in (\ref{path_nonaux}). \ For temperature $T=0.022E_{F}$ and chemical
potential $\mu=0.97E_{F}$ we apply the free field Feynman rules for the
$\psi^{2}$ correlation function. \ For this chemical potential at such low
temperatures we should see $7$ up spins and $7$ down spins in the ground state
with $\vec{P}=\vec{0}$ and $S=0$. \ A comparison of the simulation results and
the free grand canonical calculations is shown Table 2. \ We see that the free
fermion results agree to six-digit precision.%

\[
\
\genfrac{}{}{0pt}{}{%
\begin{tabular}
[c]{|c|c|c|}\hline
$n_{x}$ & Simulation results & Free grand canonical\\\hline
$0$ & $1.19629\times10^{-2}$ & $1.19629\times10^{-2}$\\\hline
$1$ & $6.10352\times10^{-3}$ & $6.10352\times10^{-3}$\\\hline
$2$ & $2.19727\times10^{-3}$ & $2.19727\times10^{-3}$\\\hline
\end{tabular}
\ }{\text{Table 2. \ Simulation results and free grand canonical calculations
for }7,7}%
\]
\bigskip

Next we turn on a weak attractive coupling $C_{\text{phys}}=-1.25\times
10^{-5}$ MeV$^{-2}$ for the same $7,7$ system with $\vec{P}=\vec{0}$ and
$S=0$. \ This coupling corresponds with a weak-coupling low-density expansion
parameter $k_{F}a_{\text{scatt}}=-0.043$. \ Using $T=0.022E_{F}$ and
$\mu=0.97E_{F}$ we use the path integral action $S(c,c^{\ast})$ to compute the
free fermion result as well as the $O\left(  k_{F}a_{\text{scatt}}\right)  $
correction to the $\psi^{2}$ correlation function. \ The $O\left(
k_{F}a_{\text{scatt}}\right)  $ term requires summing the two-particle bubble
chain shown in FIG. \ref{twotwo}. \ A comparison of the simulation results and
perturbative calculations is shown in Table 3. \ The $O\left(  k_{F}%
^{2}a_{\text{scatt}}^{2}\right)  $ correction should be of size roughly
$10^{-4}$ for $n_{x}=0$, $5\times10^{-5}$ for $n_{x}=1,$ and $2\times10^{-5}$
for $n_{x}=2$. \ We see that the simulation results and perturbative
calculations agree within deviations roughly matching the size estimates for
the $O\left(  k_{F}^{2}a_{\text{scatt}}^{2}\right)  $ term.%

\[
\
\genfrac{}{}{0pt}{}{%
\begin{tabular}
[c]{|c|c|c|}\hline
$n_{x}$ & Simulation results & Perturbative grand canonical\\\hline
$0$ & $1.3715(1)\times10^{-2}$ & $1.3775\times10^{-2}$\\\hline
$1$ & $6.999(2)\times10^{-3}$ & $7.0295\times10^{-3}$\\\hline
$2$ & $2.536(1)\times10^{-3}$ & $2.5476\times10^{-3}$\\\hline
\end{tabular}
\ }{\text{Table 3. \ Simulation results and perturbative grand canonical
calculations for }7,7}%
\]
\bigskip

To test that the code is running properly at the unitary limit, we perform
simulations for the $1,1$ system at $\vec{P}=\vec{0}$ and $S=0$ at unitarity.
$\ $We use $L=4$ and compute $G_{\psi^{2}}^{\text{bare}}(\vec{n}_{s}%
,t_{1},t_{2})$. \ We take $\vec{n}_{s}=\left\langle n_{x},0,0\right\rangle $
for $n_{x}=0,1,2$ and various numbers of time steps $n_{t_{1}}=n_{t_{2}}$.
\ Since there are only two particles with zero total momentum there should be
no dependence on the spatial coordinate $n_{x}$. \ For the numerical check we
use $M_{n_{t}}$, the transfer matrix without auxiliary fields defined in
(\ref{noauxiliary_transfer}), to compute the exact two-body transfer matrix in
the rest frame. \ The results for the lattice simulation and the exact
two-body calculation are shown in Table 4. \ We see that the simulation
results agree with the exact results up to errors the size of the estimated
stochastic noise.%

\[%
\genfrac{}{}{0pt}{}{%
\begin{tabular}
[c]{|c|c|c|c|}\hline
$n_{x}$ & $n_{t_{1}}=n_{t_{2}}=2$ & $n_{t_{1}}=n_{t_{2}}=4$ & $n_{t_{1}%
}=n_{t_{2}}=6$\\\hline
$0$ & $6.946(9)\times10^{-4}$ & $1.156(4)\times10^{-3}$ & $1.564(4)\times
10^{-3}$\\\hline
$1$ & $6.88(3)\times10^{-4}$ & $1.19(4)\times10^{-3}$ & $1.58(6)\times10^{-3}%
$\\\hline
$2$ & $6.92(1)\times10^{-4}$ & $1.147(6)\times10^{-3}$ & $1.53(3)\times
10^{-3}$\\\hline
Exact & $6.948\times10^{-4}$ & $1.153\times10^{-3}$ & $1.573\times10^{-3}%
$\\\hline
\end{tabular}
\ }{\text{Table 4: \ Simulation results and exact two-body calculation for
}1,1\text{ at unitarity}}%
\]
\bigskip

\section{Results for the renormalization constant $\Gamma$}

\begin{figure}
[ptb]
\begin{center}
\includegraphics[
height=0.8674in,
width=2.9032in
]%
{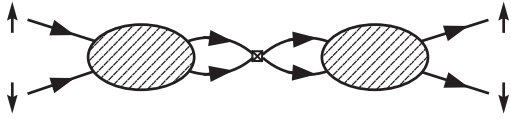}%
\caption{One insertion of the operator $a_{\downarrow}^{\dagger}(\vec
{r})a_{\uparrow}^{\dagger}(\vec{r})a_{\uparrow}(\vec{r})a_{\downarrow}(\vec
{r})$}%
\label{insert}%
\end{center}
\end{figure}

Throughout we consider systems with total momentum $\vec{P}=\vec{0}$ and total
spin $S=0$. \ The renormalization constant $\Gamma$ for the $N,N$ system is
given by%
\begin{equation}
\Gamma=\lim_{t_{1},t_{2}\rightarrow\infty}G_{\psi^{2}}^{\text{bare}}(\vec
{0},t_{1},t_{2})=\left\langle \Psi_{0}\right\vert a_{\downarrow}^{\dagger
}(\vec{0})a_{\uparrow}^{\dagger}(\vec{0})a_{\uparrow}(\vec{0})a_{\downarrow
}(\vec{0})\left\vert \Psi_{0}\right\rangle
\end{equation}
where $\left\vert \Psi_{0}\right\rangle $ is the normalized ground state.
\ $\Gamma$ gives the probability that a given lattice site has both an up-spin
and down-spin particle. \ As shown in FIG. \ref{insert}, one insertion of the
operator $a_{\downarrow}^{\dagger}(\vec{r})a_{\uparrow}^{\dagger}(\vec
{r})a_{\uparrow}(\vec{r})a_{\downarrow}(\vec{r})$ produces two extra fermion
bubbles. \ In the continuum limit each of these bubbles are proportional to
the inverse lattice spacing $a^{-1}$ plus momentum-dependent terms which are
finite as $a\rightarrow0$. \ Therefore in the continuum limit $\Gamma
_{\text{phys}}$ is proportional to $a^{-2}$ plus subleading terms proportional
to $a^{-1}$.%

The coefficient of $a^{-2}$ in $\Gamma_{\text{phys}}$ contains some useful
information about physics near the unitarity point. \ We note that
\begin{equation}
\Gamma=\left\langle \Psi_{0}\right\vert a_{\downarrow}^{\dagger}(\vec
{0})a_{\uparrow}^{\dagger}(\vec{0})a_{\uparrow}(\vec{0})a_{\downarrow}(\vec
{0})\left\vert \Psi_{0}\right\rangle \approx-\frac{1}{\alpha_{t}L^{3}}\frac
{d}{dC^{\prime}}\left\langle \Psi_{0}\right\vert M_{n_{t}}\left\vert \Psi
_{0}\right\rangle ,
\end{equation}
where $M_{n_{t}}$ is the transfer matrix without auxiliary fields defined in
(\ref{noauxiliary_transfer}), and
\begin{equation}
C^{\prime}=-\frac{\left(  e^{-C\alpha_{t}}-1\right)  \left(  1-6h\right)
^{2}}{\alpha_{t}}.
\end{equation}
We have dropped terms which are $O(a_{t})$ and vanish as the temporal lattice
spacing goes to zero. \ If $E_{0}$ is the ground state energy then%
\begin{equation}
\left\langle \Psi_{0}\right\vert M_{n_{t}}\left\vert \Psi_{0}\right\rangle
=e^{-E_{0}\alpha_{t}},
\end{equation}%
\begin{equation}
\Gamma\approx-\frac{1}{\alpha_{t}L^{3}}\frac{d}{dC^{\prime}}e^{-E_{0}%
\alpha_{t}}\approx\frac{1}{L^{3}}\frac{dE_{0}}{dC^{\prime}}. \label{gamma}%
\end{equation}
Let us parameterize the ground state\ energy per particle near the unitary
limit as an expansion in $k_{F}^{-1}a_{\text{scatt}}^{-1}$,%
\begin{equation}
\frac{E_{0}}{N+N}=\frac{3}{5}\frac{k_{F}^{2}}{2m}\left[  \xi-\xi_{1}k_{F}%
^{-1}a_{\text{scatt}}^{-1}+O(k_{F}^{-2}a_{\text{scatt}}^{-2})\right]  .
\label{near_unitary}%
\end{equation}
The renormalization condition relating $C^{\prime}$ and $a_{\text{scatt}}%
$\ gives \cite{Lee:2004qd}
\begin{equation}
\frac{d}{da_{\text{scatt}}^{-1}}=\frac{m}{4\pi}\frac{d}{dC^{\prime-1}}%
=-\frac{mC^{\prime2}}{4\pi}\frac{d}{dC^{\prime}}. \label{derivatives}%
\end{equation}
Using%
\begin{equation}
k_{F}=\frac{\left(  6\pi^{2}N\right)  ^{1/3}}{L},
\end{equation}
\ and combining (\ref{gamma}), (\ref{near_unitary}), (\ref{derivatives}) we
find%
\begin{equation}
\xi_{1}=-\frac{5m}{3Nk_{F}}\frac{dE_{0}}{da_{\text{scatt}}^{-1}}=\frac
{5m^{2}C^{\prime2}L^{4}}{12\pi\left(  6\pi^{2}\right)  ^{1/3}N^{4/3}}\Gamma.
\label{xsi1}%
\end{equation}
Since $C_{\text{phys}}^{\prime}$ is proportional to the lattice spacing $a$,
we deduce that the leading divergence of $\Gamma_{\text{phys}}$ is
proportional to $a^{-2}$ as predicted before. \ This anomalous dimensional
scaling can also be seen from the $L^{-4}$ dependence of $\Gamma$ rather than
the naive $L^{-6}$ scaling expected for an operator which is the square of a
local density.

We determine $\Gamma$ by fitting $G_{\psi^{2}}^{\text{bare}}(\vec{0},t/2,t/2)$
at large $E_{F}t$ to the asymptotic form%
\begin{equation}
G_{\psi^{2}}^{\text{bare}}(\vec{0},t/2,t/2)\approx\Gamma-be^{-\eta\cdot
E_{F}t}\text{.}%
\end{equation}
\ The value of $\Gamma$\ is then used to determine $\xi_{1}$. \ We show the
results for $N=3,5,7,9,11,13$ and $L=4,5,6$ in Table 5. \ We have extrapolated
linearly in $L^{-1}$ as $L\rightarrow\infty$ to remove the subleading $a^{-1}$
dependence in $G_{\psi^{2}}^{\text{bare}}(\vec{0},t/2,t/2)$. \ If there are no
significant changes for $N>13$ we estimate that in the limit $N\rightarrow
\infty$, $\xi_{1}=1.0(1).$%

\[%
\genfrac{}{}{0pt}{}{%
\begin{tabular}
[c]{|c|c|c|c|c|c|c|}\hline
$L$ & $3,3$ & $5,5$ & $7,7$ & $9,9$ & $11,11$ & $13,13$\\\hline
$4$ & $0.696(2)$ & $0.647(2)$ & $0.597(2)$ & $0.595(2)$ & $-$ & $-$\\\hline
$5$ & $0.77(1)$ & $0.719(5)$ & $0.662(3)$ & $0.661(2)$ & $0.652(2)$ &
$0.639(2)$\\\hline
$6$ & $0.84(3)$ & $0.785(4)$ & $0.69(1)$ & $0.711(4)$ & $0.71(1)$ &
$0.70(1)$\\\hline
$\infty$ & $1.08(4)$ & $1.05(4)$ & $0.91(4)$ & $0.93(4)$ & $1.00(5)$ &
$1.01(5)$\\\hline
\end{tabular}
\ }{\text{Table 5: \ Results for }\xi_{1}}%
\]
\bigskip

\begin{figure}
[ptb]
\begin{center}
\includegraphics[
height=2.5244in,
width=3.5276in
]%
{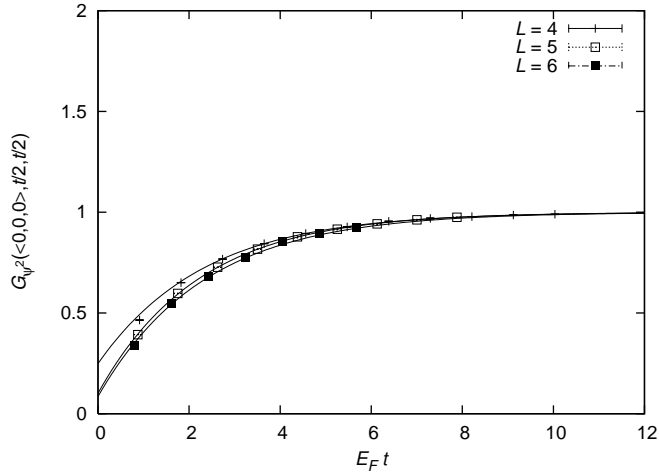}%
\caption{A comparison of $G_{\psi^{2}}(\vec{0},t/2,t/2)$ and the asymptotic
fit $1-\frac{b}{\Gamma}e^{-\eta\cdot E_{F}t}$ for the $9,9$ system with
$L=4,5,6.$}%
\label{n9peak_s}%
\end{center}
\end{figure}

In FIG. \ref{n9peak_s} we show a comparison of the renormalized correlation
function%
\begin{equation}
G_{\psi^{2}}(\vec{0},t/2,t/2)=\frac{1}{\Gamma}G_{\psi^{2}}^{\text{bare}}%
(\vec{0},t/2,t/2)
\end{equation}
and the asymptotic fit%
\begin{equation}
1-\frac{b}{\Gamma}e^{-\eta\cdot E_{F}t}%
\end{equation}
for the $9,9$ system and $L=4,5,6$. \ In the unitary limit we expect agreement
for different values of $L$ when plotted as functions of $E_{F}t$, and this
appear to be the case.%

The energy $\eta\cdot E_{F}$ characterizing the exponential decay of the
transient signal in $G_{\psi^{2}}(\vec{0},t/2,t/2)$ is similar to the energy
$\delta\cdot E_{F}$ measured in the function $\xi_{N,N}(t)$ described in
\cite{Lee:2005fk}. \ This might perhaps be a threshold for a certain type of
excitation or a maximum in the overlap of $\left\vert \Psi_{0}^{\text{free}%
}\right\rangle $ with the spectral density projection operator. \ The answer
is not clear. \ In the next section we find another excitation with the same
quantum numbers but much lower energy. \ This suggests that our fit with only
one exponential time constant may not be a reliable method to determine the
higher excitation energy, and a multistate analysis should be used. \ The
interpretation of the apparent energy scale $\eta\cdot E_{F}$ will require
further study.

\section{Results for the $t$-dependent profile of $G_{\psi^{2}}$}

\begin{figure}
[ptb]
\begin{center}
\includegraphics[
height=3.026in,
width=3.3875in
]%
{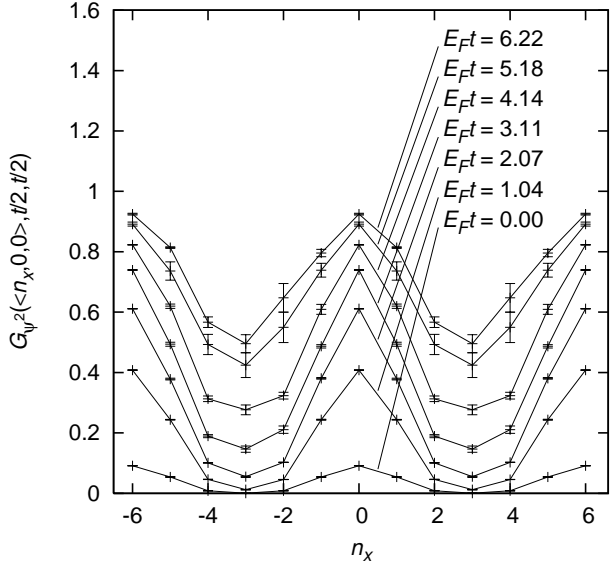}%
\caption{The renormalized correlation function $G_{\psi^{2}}(\vec{n}%
_{s},t/2,t/2)$ for the $13,13$ system with $L=6$, $\vec{n}_{s}=\left\langle
n_{x},0,0\right\rangle $, and $E_{F}t$ between $0$ and $6.22$.}%
\label{stroben13_x}%
\end{center}
\end{figure}

We study the dependence of $G_{\psi^{2}}$ on the spatial separation $\vec
{n}_{s}$ and Euclidean projection time $t$. \ In all cases we consider systems
with total momentum $\vec{P}=\vec{0}$ and total spin $S=0$. \ In FIG.
\ref{stroben13_x} we show the renormalized correlation function $G_{\psi^{2}%
}(\vec{n}_{s},t/2,t/2)$ for the $13,13$ system at unitarity with $L=6$,
$\vec{n}_{s}=\left\langle n_{x},0,0\right\rangle $, and $E_{F}t$ between $0$
and $6.22$. \ In order to make the periodicity of the lattice visually clear
we show two full lattice lengths.%

There are several features in this plot which indicate some interesting
physics. \ First of all $\psi^{2}$ appears to have long-range order. \ With a
total of $26$ particles we cannot probe distances far beyond $k_{F}^{-1}$.
\ But for large $E_{F}t$ we do find that $G_{\psi^{2}}(\vec{n}_{s},t/2,t/2)$
averaged over the entire lattice is greater than $0.5$. \ We recall
$G_{\psi^{2}}(\vec{n}_{s},t/2,t/2)$ is normalized so that the peak value at
$\vec{n}_{s}=\vec{0}$ is $1$ as $E_{F}t\rightarrow\infty$. \ The second point
of interest is that for $\vec{n}_{s}\neq\vec{0}$ the transient signal has a
very long time constant. \ This long constant time is not apparent at $\vec
{n}_{s}=\vec{0}$. \ There the transient signal has a shorter time constant,
$(\eta\cdot E_{F})^{-1}$ in the notation of the previous section. \ The slow
transient signal corresponds with a much lower energy scale and appears to
have an approximate $\cos(2\pi n_{x}/L)-1$ spatial dependence.

\begin{figure}
[ptb]
\begin{center}
\includegraphics[
height=4.2186in,
width=4.2272in
]%
{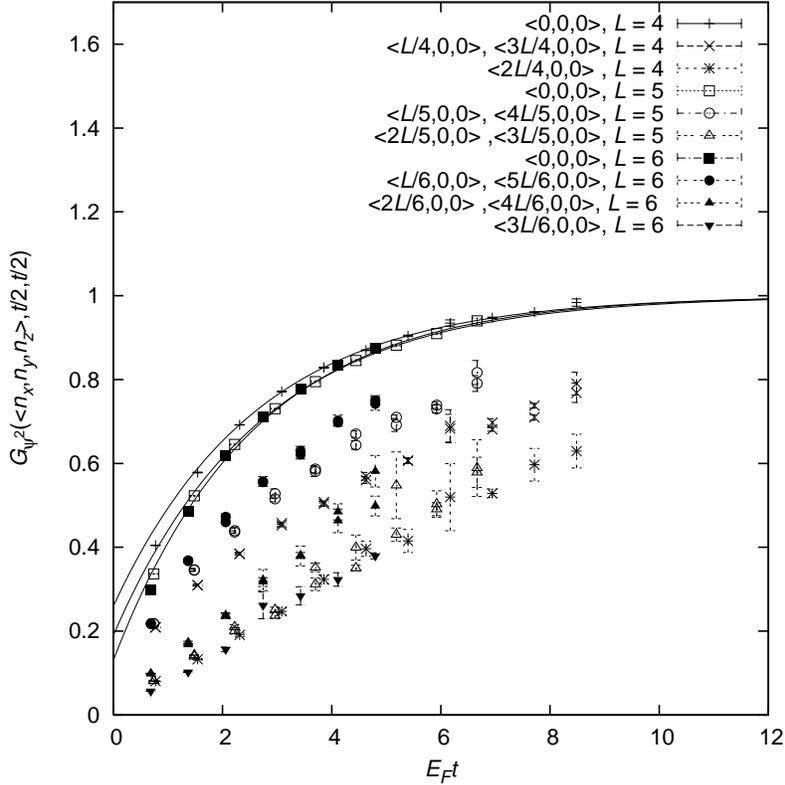}%
\caption{The renormalized correlation function $G_{\psi^{2}}(\vec{n}%
_{s},t/2,t/2)$ versus $E_{F}t$ for the $7,7$ system with $L=4,5,6$ and
$\vec{n}_{s}=\left\langle n_{x},0,0\right\rangle $. \ For visual clarity
datapoints with errorbars exceeding $0.1$ are not shown.}%
\label{profilen7_x}%
\end{center}
\end{figure}

The time constant is easier to see if we plot $G_{\psi^{2}}(\vec{n}%
_{s},t/2,t/2)$ as a function of $E_{F}t$. \ In FIG. \ref{profilen7_x} we show
the renormalized correlation function $G_{\psi^{2}}(\vec{n}_{s},t/2,t/2)$
versus $E_{F}t$ for the $7,7$ system with $L=4,5,6$ and $\vec{n}%
_{s}=\left\langle n_{x},0,0\right\rangle $. \ We can see quite clearly the
fast exponential tail of the transient signal for $\vec{n}_{s}=\vec{0}$ and
the slow exponential tail for $\vec{n}_{s}\neq\vec{0}$. \ In the unitary
limit, $G_{\psi^{2}}(\vec{n}_{s},t/2,t/2)$ as a function of $E_{F}t$ for
different $L$ should agree for the same ratio $\vec{n}_{s}/L$. \ The data in
FIG. \ref{profilen7_x} shows quite clearly this unitary limit scaling.%

\begin{figure}
[ptb]
\begin{center}
\includegraphics[
height=4.2177in,
width=4.2272in
]%
{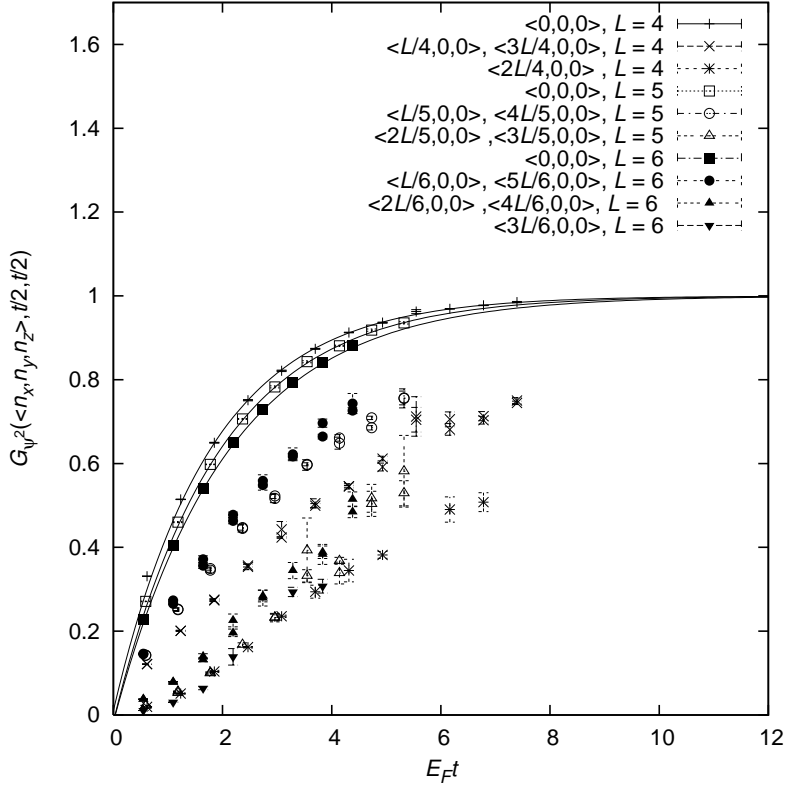}%
\caption{The renormalized correlation function $G_{\psi^{2}}(\vec{n}%
_{s},t/2,t/2)$ versus $E_{F}t$ for the $5,5$ system with $L=4,5,6$ and
$\vec{n}_{s}=\left\langle n_{x},0,0\right\rangle $. \ For visual clarity
datapoints with errorbars exceeding $0.1$ are not shown.}%
\label{profilen5_x}%
\end{center}
\end{figure}

In FIG. \ref{profilen5_x} the renormalized correlation function $G_{\psi^{2}%
}(\vec{n}_{s},t/2,t/2)$ is shown for the $5,5$ system with $L=4,5,6$ and
$\vec{n}_{s}=\left\langle n_{x},0,0\right\rangle $. \ We again see the fast
exponential tail of the transient signal at $\vec{n}_{s}=\vec{0}$ and the slow
exponential tail for $\vec{n}_{s}\neq\vec{0}$. \ There is good agreement for
different values of $L$ with the same $\vec{n}_{s}/L$.%

\begin{figure}
[ptbptb]
\begin{center}
\includegraphics[
height=4.2177in,
width=4.2272in
]%
{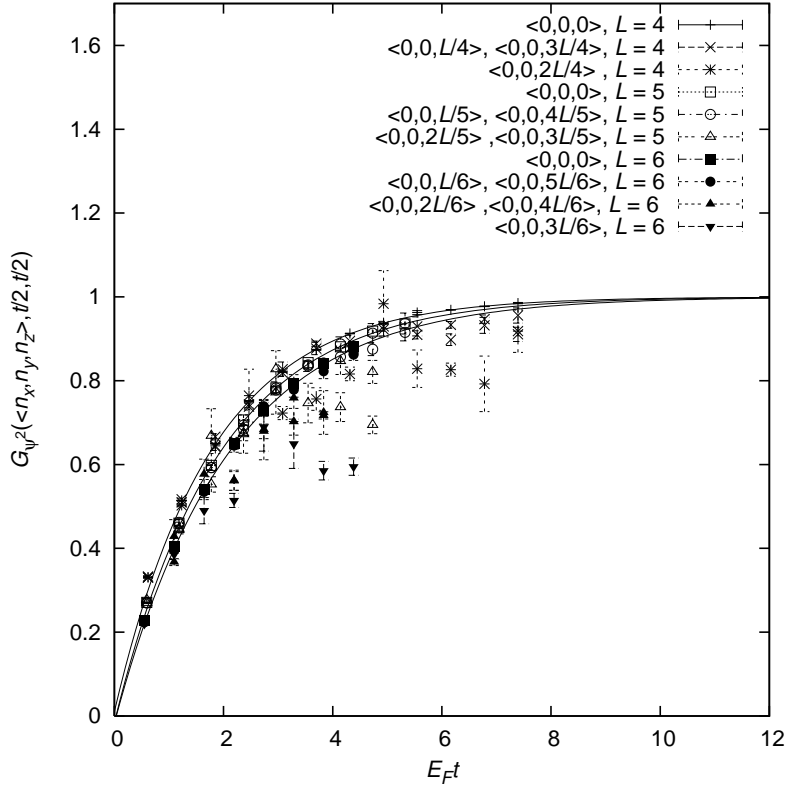}%
\caption{The renormalized correlation function $G_{\psi^{2}}(\vec{n}%
_{s},t/2,t/2)$ versus $E_{F}t$ for the $5,5$ system with $L=4,5,6$ and
$\vec{n}_{s}=\left\langle 0,0,n_{z}\right\rangle $. \ For visual clarity
datapoints with errorbars exceeding $0.1$ are not shown.}%
\label{profilen5_z}%
\end{center}
\end{figure}

In FIG. \ref{profilen5_z} $G_{\psi^{2}}(\vec{n}_{s},t/2,t/2)$ is shown for the
same $5,5$ system with $L=4,5,6,$ but this time along the $z$-axis, $\vec
{n}_{s}=\left\langle 0,0,n_{z}\right\rangle $. \ In this case the long time
constant previously seen for $\vec{n}_{s}\neq\vec{0}$ is no longer visible.
\ The distinction between the $x$- and $z$-axes can be explained by the
non-$SO(3,\mathbb{Z})$ invariant momentum filling for $\left\vert \Psi
_{0}^{\text{free}}\right\rangle $. \ As shown in Table 1, $\left\vert \Psi
_{0}^{\text{free}}\right\rangle $ contains the momentum states $\left\langle
\frac{2\pi}{L},0,0\right\rangle $ and $\left\langle -\frac{2\pi}%
{L},0,0\right\rangle $ but not the momentum states $\left\langle
0,0,\frac{2\pi}{L}\right\rangle $ and $\left\langle 0,0,-\frac{2\pi}%
{L}\right\rangle $. \ This produces an overlap with some low-energy excitation
with an approximate $\cos(2\pi n_{x}/L)-1$ profile in $G_{\psi^{2}}(\vec
{n}_{s},t/2,t/2)$ but a much weaker overlap if the same excitation is aligned
along the $z$-axis.%

\section{Results for the lowest excitation energy}

We now measure the energy of the lowest excitation. \ We first construct
combinations of $G_{\psi^{2}}(\vec{n}_{s},t_{1},t_{2})$ which maximize the
signal to noise ratio. \ For $N=3$ the low-energy excitation does not couple
strongly to $G_{\psi^{2}}(\vec{n}_{s},t_{1},t_{2})$ for $\vec{n}_{s}$ along
the $y$- and $z$-axes. \ Therefore we use only data along the $x$-axis. \ We
define%
\begin{align}
&  \delta_{x}^{2}G_{\psi^{2}}(t_{1},t_{2})\nonumber\label{deltax}\\
&  =L^{2}\left\{  G_{\psi^{2}}(\vec{0},t_{1},t_{2})-\frac{1}{2}\left[
G_{\psi^{2}}(\left\langle 1,0,0\right\rangle ,t_{1},t_{2})+G_{\psi^{2}%
}(\left\langle -1,0,0\right\rangle ,t_{1},t_{2})\right]  \right\}  .
\end{align}
For $N=5$ the coupling is weak only along the $z$-axis and so we define%
\begin{align}
&  \delta_{xy}^{2}G_{\psi^{2}}(t_{1},t_{2})\nonumber\label{deltaxy}\\
&  =L^{2}\left\{  G_{\psi^{2}}(\vec{0},t_{1},t_{2})-\frac{1}{4}\left[
\begin{array}
[c]{c}%
G_{\psi^{2}}(\left\langle 1,0,0\right\rangle ,t_{1},t_{2})+G_{\psi^{2}%
}(\left\langle -1,0,0\right\rangle ,t_{1},t_{2})\\
+G_{\psi^{2}}(\left\langle 0,1,0\right\rangle ,t_{1},t_{2})+G_{\psi^{2}%
}(\left\langle 0,-1,0\right\rangle ,t_{1},t_{2})
\end{array}
\right]  \right\}  .
\end{align}
For $N\geq7$ we define%
\begin{align}
&  \delta_{xyz}^{2}G_{\psi^{2}}(t_{1},t_{2})\nonumber\label{deltaxyz}\\
&  =L^{2}\left\{  G_{\psi^{2}}(\vec{0},t_{1},t_{2})-\frac{1}{6}\left[
\begin{array}
[c]{c}%
G_{\psi^{2}}(\left\langle 1,0,0\right\rangle ,t_{1},t_{2})+G_{\psi^{2}%
}(\left\langle -1,0,0\right\rangle ,t_{1},t_{2})\\
+G_{\psi^{2}}(\left\langle 0,1,0\right\rangle ,t_{1},t_{2})+G_{\psi^{2}%
}(\left\langle 0,-1,0\right\rangle ,t_{1},t_{2})\\
+G_{\psi^{2}}(\left\langle 0,0,1\right\rangle ,t_{1},t_{2})+G_{\psi^{2}%
}(\left\langle 0,0,-1\right\rangle ,t_{1},t_{2})
\end{array}
\right]  \right\}  .
\end{align}

\begin{figure}
[ptb]
\begin{center}
\includegraphics[
height=1.2021in,
width=4.5913in
]%
{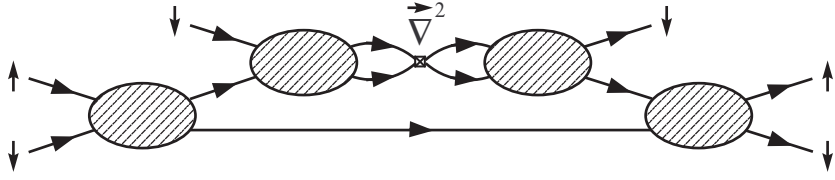}%
\caption{One insertion of the operator $\frac{1}{\Gamma}a_{\downarrow
}^{\dagger}(\vec{r})a_{\uparrow}^{\dagger}(\vec{r})\vec{\nabla}^{2}\left[
a_{\uparrow}(\vec{r})a_{\downarrow}(\vec{r})\right]  .$}%
\label{insertderivloop}%
\end{center}
\end{figure}

If no additional ultraviolet renormalization is required, then in the
continuum limit we have%
\begin{align}
\delta_{x}^{2}G_{\psi^{2}}(t_{1},t_{2})  &  \propto-\partial_{x}^{2}%
G_{\psi^{2}}(\vec{0},t_{1},t_{2}),\\
\delta_{xy}^{2}G_{\psi^{2}}(t_{1},t_{2})  &  \propto-\left(  \partial_{x}%
^{2}+\partial_{y}^{2}\right)  G_{\psi^{2}}(\vec{0},t_{1},t_{2}),\\
\delta_{xyz}^{2}G_{\psi^{2}}(t_{1},t_{2})  &  \propto-\left(  \partial_{x}%
^{2}+\partial_{y}^{2}+\partial_{z}^{2}\right)  G_{\psi^{2}}(\vec{0}%
,t_{1},t_{2}).
\end{align}
To prove that no additional renormalization is needed it suffices to show that
one insertion of the operator%
\begin{equation}
\frac{1}{\Gamma}a_{\downarrow}^{\dagger}(\vec{r})a_{\uparrow}^{\dagger}%
(\vec{r})\vec{\nabla}^{2}\left[  a_{\uparrow}(\vec{r})a_{\downarrow}(\vec
{r})\right]
\end{equation}
is finite. \ Just as we found for the insertion of $a_{\downarrow}^{\dagger
}(\vec{r})a_{\uparrow}^{\dagger}(\vec{r})a_{\uparrow}(\vec{r})a_{\downarrow
}(\vec{r})$, the factor of $\Gamma^{-1}$ takes care of the divergences from
the two extra fermion bubbles. \ In the unitary limit the two-particle Green's
function has the form%
\begin{equation}
G_{2}(p_{0},\vec{p})\propto\frac{1}{m\sqrt{mp_{0}-\frac{\vec{p}^{2}}{4}}},
\end{equation}
where $p_{0}$ is the total energy and $\vec{p}$ is the total momentum of the
two fermions. \ Consider now a diagram such as the one shown in FIG.
\ref{insertderivloop}.%
Let $p_{0}$ be the energy and $\vec{p}$ be the momentum of the internal loop.
\ We cutoff the momentum at $\Lambda=\pi a^{-1}$ and cutoff the energy at
$\Lambda^{2}/m$. If we now count powers of $\Lambda$ we get $\Lambda^{5}$ from
$dp_{0}d^{3}\vec{p}$, $\Lambda^{-2}$ from the two Green's functions,
$\Lambda^{-6}$ from the three fermion propagators, and $\Lambda^{2}$ from the
$\vec{\triangledown}^{2}$. \ The final result is $\Lambda^{-1}$ and so the
integral is finite. \ In this way one can show that all diagrams involving one
insertion of
\begin{equation}
\frac{1}{\Gamma}a_{\downarrow}^{\dagger}(\vec{r})a_{\uparrow}^{\dagger}%
(\vec{r})\vec{\nabla}^{2}\left[  a_{\uparrow}(\vec{r})a_{\downarrow}(\vec
{r})\right]
\end{equation}
are ultraviolet finite. \ This means that in the unitary limit each of the
functions $\delta_{x}^{2}G_{\psi^{2}}(t_{1},t_{2})$, $\delta_{xy}^{2}%
G_{\psi^{2}}(t_{1},t_{2})$, $\delta_{xyz}^{2}G_{\psi^{2}}(t_{1},t_{2})$ should
be independent of $L$ when considered as functions of $E_{F}t_{1}$ and
$E_{F}t_{2}$.

\begin{figure}
[ptb]
\begin{center}
\includegraphics[
height=3.0234in,
width=4.2272in
]%
{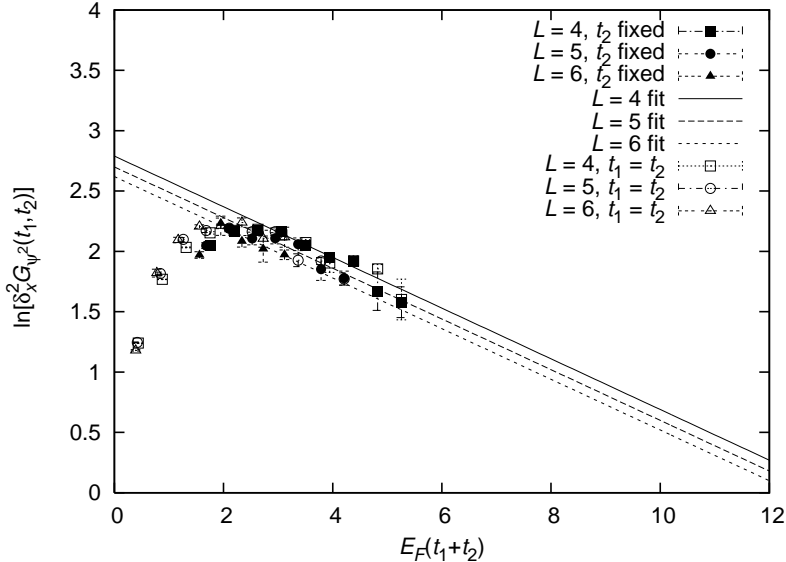}%
\caption{Plot of $\ln\left[  \delta_{x}^{2}G_{\psi^{2}}(t_{1},t_{2})\right]  $
for the $3,3$ system with $L=4,5,6$. \ We show data for $t_{1}=t_{2}$ and
$t_{2}$ fixed at $E_{F}t_{2}=1.2$.}%
\label{logn3_asym}%
\end{center}
\end{figure}

In FIG. \ref{logn3_asym} we plot $\ln\left[  \delta_{x}^{2}G_{\psi^{2}}%
(t_{1},t_{2})\right]  $ for the $3,3$ system with $L=4,5,6$. \ We have
produced data for $t_{1}=t_{2}$ and data for $t_{2}$ fixed at $E_{F}t_{2}%
=1.2$. \ The agreement for different values of $L$ provides a consistency
check of unitary limit scaling. \ In the continuum language we are calculating
the logarithm of $-\partial_{x}^{2}G_{\psi^{2}}(\vec{0},t_{1},t_{2})$. \ Using
the asymptotic expansion (\ref{G_asymp}) we have%
\begin{align}
-\partial_{x}^{2}G_{\psi^{2}}(\vec{0},t_{1},t_{2})  &  =-\partial_{x}%
^{2}A_{00}(\vec{0})-\partial_{x}^{2}A_{01}(\vec{0})e^{-(E_{1}-E_{0})t_{2}%
}\nonumber\\
&  -\partial_{x}^{2}A_{10}(\vec{0})e^{-(E_{1}-E_{0})t_{1}}-\partial_{x}%
^{2}A_{11}(\vec{0})e^{-(E_{1}-E_{0})(t_{1}+t_{2})}+\cdots.
\end{align}
The fixed $t_{2}$ data as $t_{1}\rightarrow\infty$ is useful in extracting
$E_{1}-E_{0}$ since the time dependence must be proportional to $e^{-(E_{1}%
-E_{0})t_{1}}$ plus a constant. \ We see from the plot that $\ln\left[
\delta_{x}^{2}G_{\psi^{2}}(t_{1},t_{2})\right]  $ is nearly a straight line
for large $t_{1}$. $\ $This indicates a small asymptotic value at
$t_{1}=\infty$, and a nearly pure exponential signal in this time window. \ We
have fitted a straight line to determine the slope of $\ln\left[  \delta
_{x}^{2}G_{\psi^{2}}(t_{1},t_{2})\right]  $ with respect to $E_{F}t_{1}$ for
the fixed $t_{2}$ data. \ We do the fit with a common slope for $L=4,5,6$ but
possibly different intercepts for the three values of $L.$ \ While we are only
fitting the fixed $t_{2}$ data, we see quite clearly the same slope appears in
the $t_{1}=t_{2}$ data.%

\begin{figure}
[ptb]
\begin{center}
\includegraphics[
height=3.0234in,
width=4.2272in
]%
{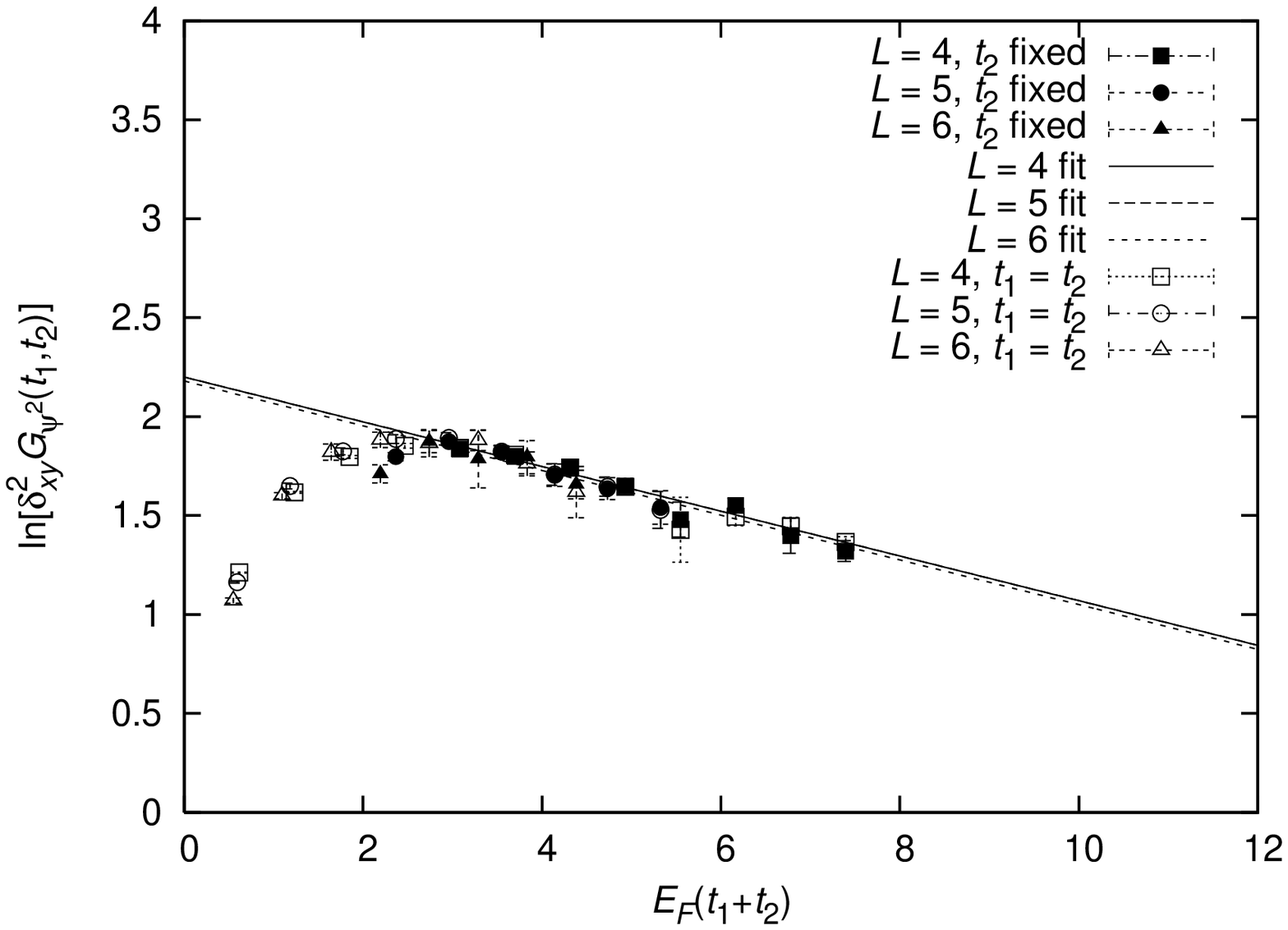}%
\caption{Plot of $\ln\left[  \delta_{xy}^{2}G_{\psi^{2}}(t_{1},t_{2})\right]
$ for the $5,5$ system with $L=4,5,6$. \ We show data for $t_{1}=t_{2}$ and
$t_{2}$ fixed at $E_{F}t_{2}=1.6.$}%
\label{logn5_asym}%
\end{center}
\end{figure}

\begin{figure}
[ptb]
\begin{center}
\includegraphics[
height=3.0234in,
width=4.2272in
]%
{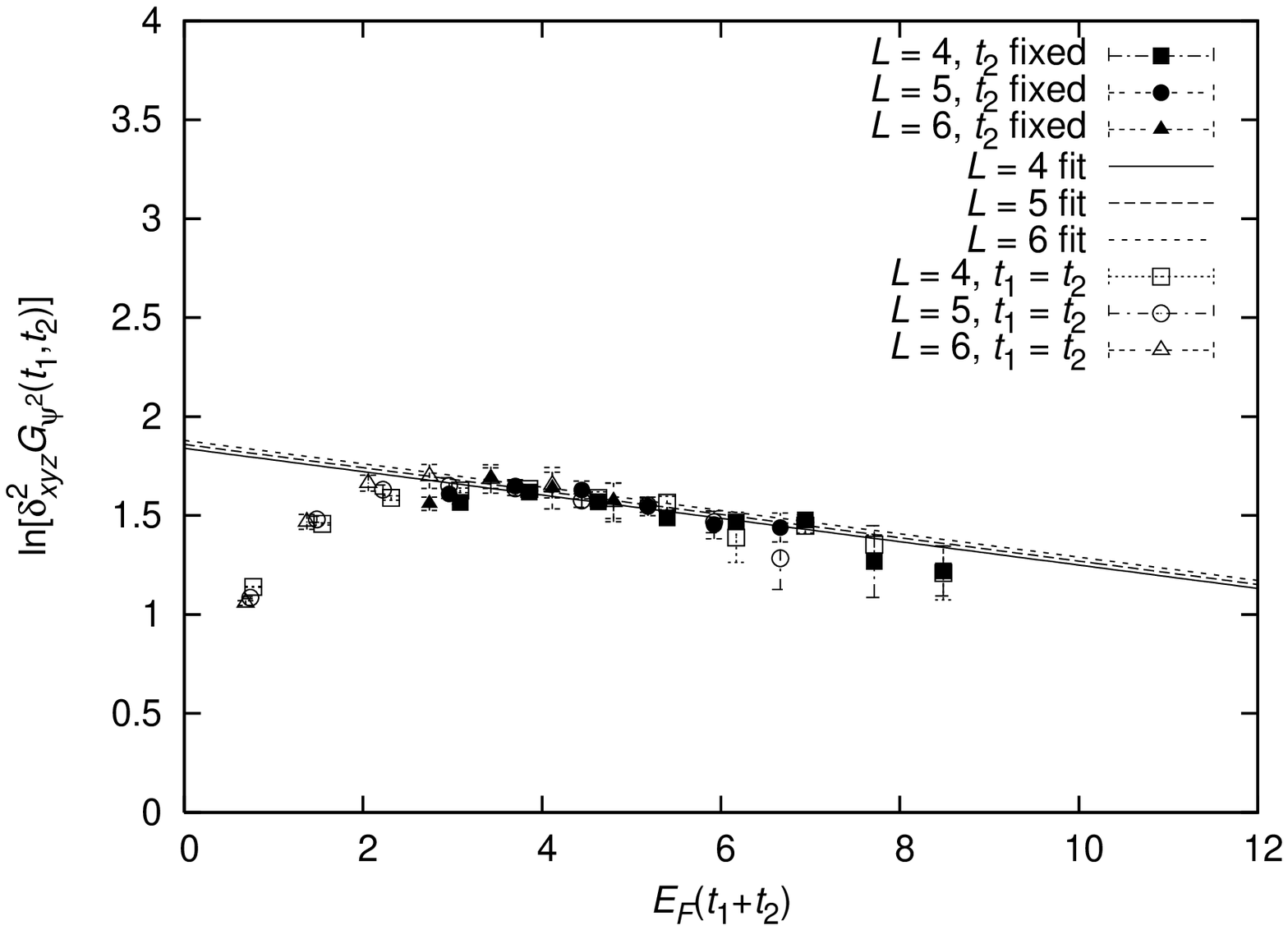}%
\caption{Plot of $\ln\left[  \delta_{xyz}^{2}G_{\psi^{2}}(t_{1},t_{2})\right]
$ for the $7,7$ system with $L=4,5,6$. \ We show data for $t_{1}=t_{2}$ and
$t_{2}$ fixed at $E_{F}t_{2}=2.0.$}%
\label{logn7_asym}%
\end{center}
\end{figure}

\begin{figure}
[ptb]
\begin{center}
\includegraphics[
height=3.0234in,
width=4.2272in
]%
{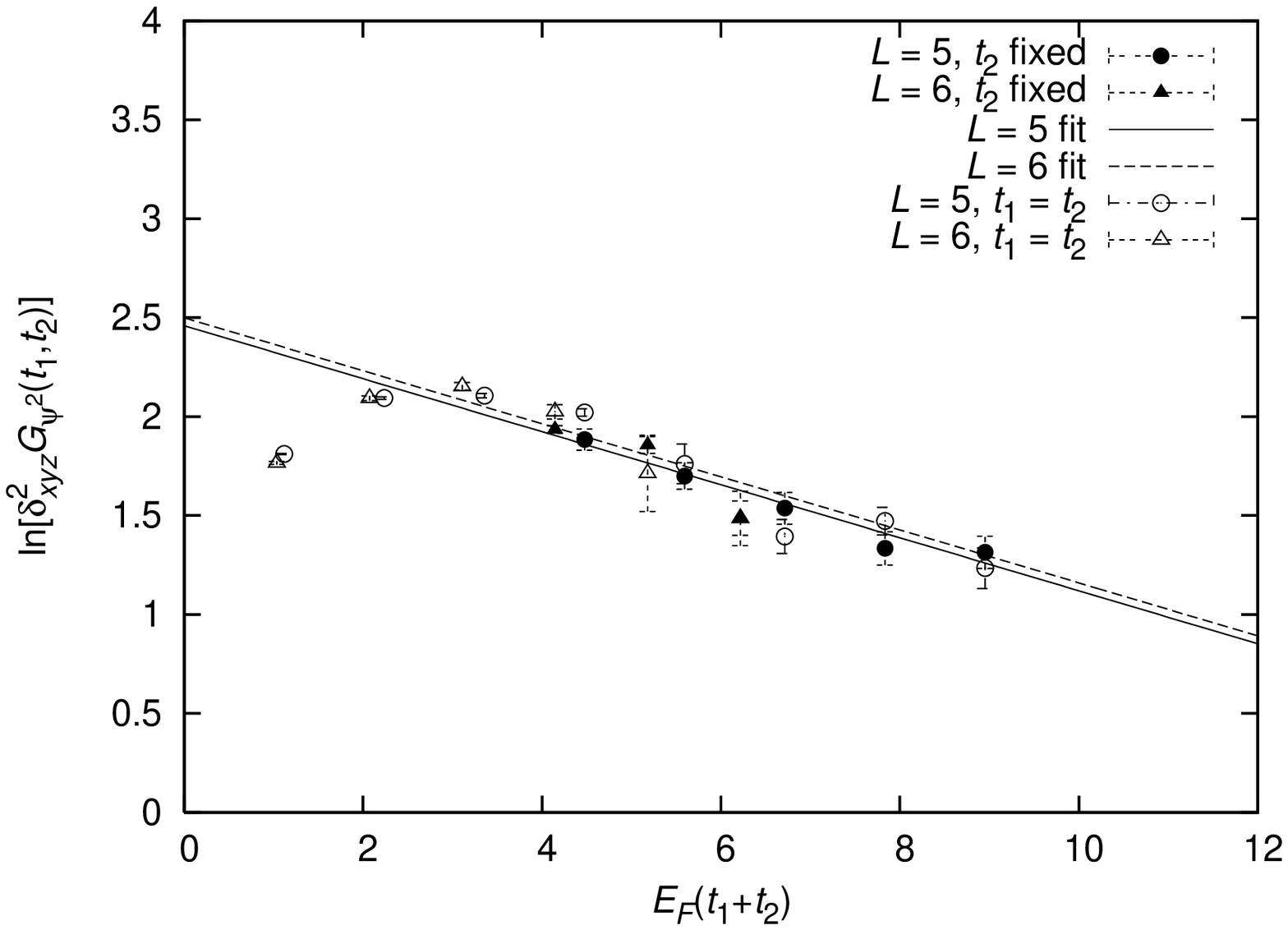}%
\caption{Plot of $\ln\left[  \delta_{xyz}^{2}G_{\psi^{2}}(t_{1},t_{2})\right]
$ for the $13,13$ system with $L=5,6$. \ We show data for $t_{1}=t_{2}$ and
$t_{2}$ fixed at $E_{F}t_{2}=3.0.$}%
\label{logn13_asym}%
\end{center}
\end{figure}

In FIG. \ref{logn5_asym}\ we plot $\ln\left[  \delta_{xy}^{2}G_{\psi^{2}%
}(t_{1},t_{2})\right]  $ for the $5,5$ system with $L=4,5,6$. \ We show data
for $t_{1}=t_{2}$ and $t_{2}$ fixed at $E_{F}t_{2}=1.6.$ \ We do the same
linear fits for the $t_{2}$ fixed data in order to extract $E_{1}-E_{0}$. \ In
FIG. \ref{logn7_asym} we plot $\ln\left[  \delta_{xyz}^{2}G_{\psi^{2}}%
(t_{1},t_{2})\right]  $ for the $7,7$ system with $L=4,5,6.$ \ In FIG.
\ref{logn13_asym} we plot $\ln\left[  \delta_{xyz}^{2}G_{\psi^{2}}(t_{1}%
,t_{2})\right]  $ for the $13,13$ system with $L=5,6.$%

%

%

In all cases we find agreement for different values of $L$ as predicted by
unitary limit scaling. \ In all cases we also find agreement between the slope
of the fixed $t_{2}$ data and the $t_{1}=t_{2}$ data when plotted as functions
of $E_{F}(t_{1}+t_{2}$). \ This implies that the $e^{-(E_{1}-E_{0})t_{1}}$ and
$e^{-(E_{1}-E_{0})t_{2}}$ terms are small so that%
\begin{equation}
-\vec{\nabla}^{2}G_{\psi^{2}}(\vec{0},t_{1},t_{2})\approx-\vec{\nabla}%
^{2}A_{00}(\vec{0})-\vec{\nabla}^{2}A_{11}(\vec{0})e^{-(E_{1}-E_{0}%
)(t_{1}+t_{2})},
\end{equation}
with $\vec{\nabla}^{2}$ replaced by $\partial_{x}^{2}$ for the $3,3$ system
and $\partial_{x}^{2}+\partial_{y}^{2}$ for the $5,5$ system$.$

\begin{figure}
[ptb]
\begin{center}
\includegraphics[
height=3.0234in,
width=4.2272in
]%
{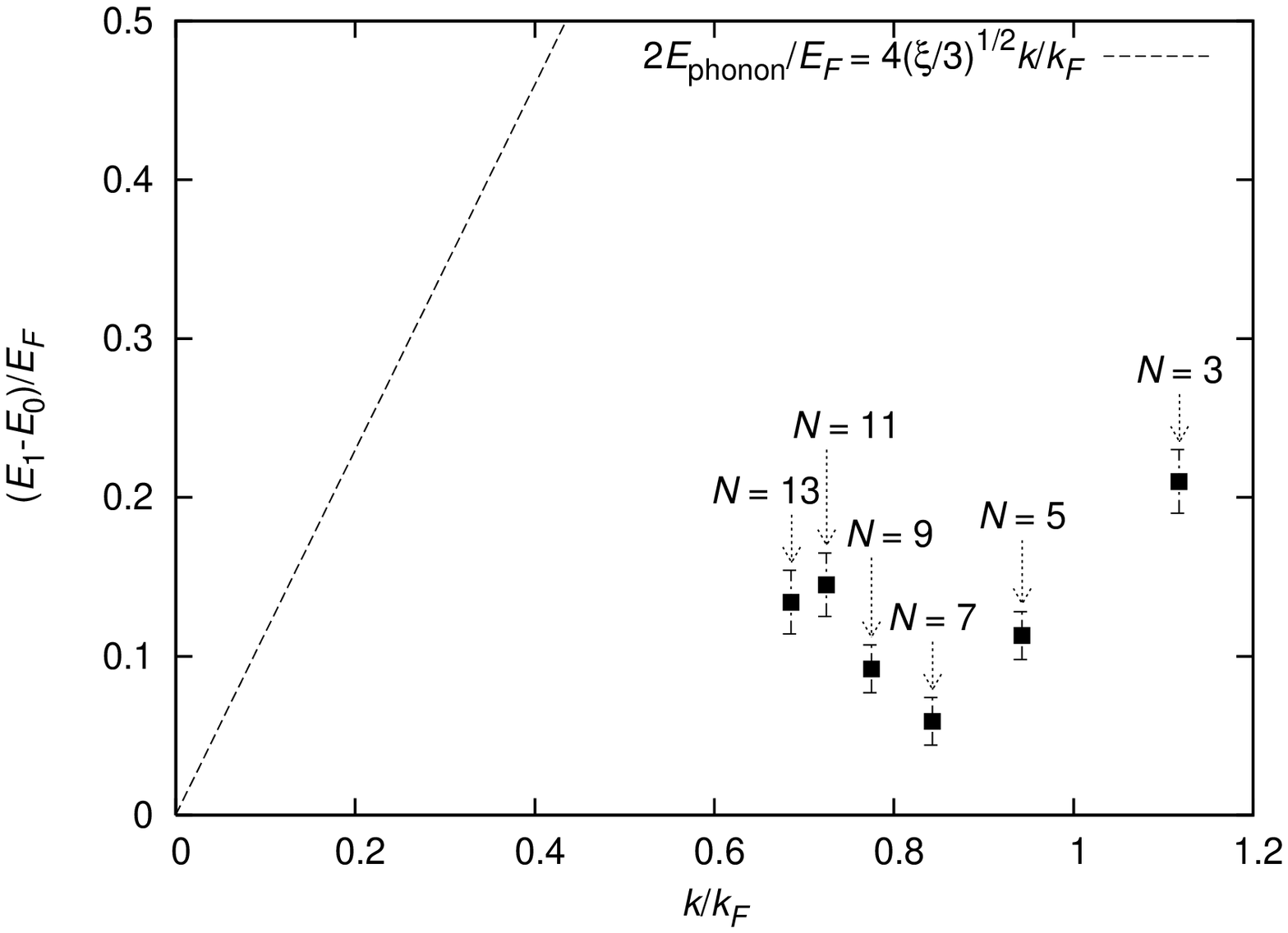}%
\caption{ Plot of $(E_{1}-E_{0})/E_{F}$ versus momentum $k/k_{F}$ of the
unknown consituents. \ For comparison we show the linear part of the
two-phonon dispersion relation at unitarity.}%
\label{dispersion}%
\end{center}
\end{figure}

For the moment let us assume that the excitation at energy $E_{1}$ can be
described as two unknown constituents moving in opposite directions with
momentum $\frac{2\pi}{L}$, the minimum nonzero momentum possible on the
lattice. \ This interpretation of the excitation is probably an
oversimplification, but it serves as a reasonable starting point to compare
with known excitations such as pairs of phonons, quasiparticles, or rotons.
\ In FIG. \ref{dispersion} we plot $(E_{1}-E_{0})/E_{F}$ versus momentum
$k/k_{F}$ of the unknown constituents for $N=3,5,7,9,11,13$. \ For comparison
we show the linear part of the two-phonon dispersion relation at unitarity.
\ We use the result \cite{Son:2005gen}%
\begin{equation}
c_{s}=\frac{k_{F}}{m}\sqrt{\frac{\xi}{3}}%
\end{equation}
for the speed of sound and use the value $\xi=0.25(3)$ reported in
\cite{Lee:2005fk}. \ We see that the minimum in $(E_{1}-E_{0})/E_{F}$ at
$k\approx0.8k_{F}$ falls well below the linear extrapolation for two phonons
at $k\approx0.8k_{F}$. \ This appears to rule out a two phonon interpretation.%

Since $0.8k_{F}$ is close to $k_{F}$, another reasonable interpretation is
that the excitation consists of two fermionic quasiparticles. \ However here
we encounter a similar problem. \ The energy $E_{1}-E_{0}$ is far below
current estimates of $2\Delta$, where $\Delta$ is the even-odd staggering in
the ground state energy. \ Fixed-node Green's function Monte Carlo simulations
get a value $\Delta=0.57(3)E_{F}$ for $12-20$ particles in a periodic cube
\cite{Chang:2004PRA} and $\Delta=0.50(3)E_{F}$ for $54-66$ particles
\cite{Carlson:2005xy}. \ Our own lattice simulations for $\Delta$ are in
progress, but there is no evidence for $\Delta$ being nearly an order of
magnitude lower than the fixed-node Monte Carlo results. \ It could be that
the quasiparticles are interacting strongly, but some mechanism would be
needed to explain the significant lowering in energy.

A third possibility we consider is that the excitation is a pair of rotons
moving in opposite directions. \ The shape of the dispersion curve in Fig.
\ref{dispersion} is similar to the phonon-roton spectrum in superfluid $^{4}%
$He. \ Landau \cite{Landau:1941a,Landau:1947a} first predicted the existence
of a roton minimum in superfluid $^{4}$He,%
\begin{equation}
E(k)\approx\Delta_{R}+\frac{(k-k_{R})^{2}}{2\mu_{R}},
\end{equation}
and\ Feynman gave a quantum-mechanical description of a roton as an excitation
whose wavelength is resonant with the local spatial correlations of identical
Bose particles \cite{Feynman:1953b,Feynman:1954a,Feynman:1956a}. \ We briefly
summarize the argument below.

Let $\phi_{0}\left(  \vec{r}_{1},\cdots,\vec{r}_{N}\right)  $ be the ground
state wavefunction for an interacting system of $N$ identical bosons with
energy $E_{0}$. \ We construct a trial state $\psi_{\vec{k}}\left(  \vec
{r}_{1},\cdots,\vec{r}_{N}\right)  $ with momentum $\vec{k}$ defined as%
\begin{equation}
\psi_{\vec{k}}\left(  \vec{r}_{1},\cdots,\vec{r}_{N}\right)  =\sum_{i=1}%
^{N}e^{i\vec{k}\cdot\vec{r}_{i}}\times\phi_{0}\left(  \vec{r}_{1},\cdots
,\vec{r}_{N}\right)  .
\end{equation}
The static structure factor $S(k)$ for the ground state can be written in
terms of the square of the norm of $\psi_{\vec{k}}$,%
\begin{align}
N\times S(k)  &  =%
{\displaystyle\int}
d^{3N}\vec{r}\;\sum_{l,m=1}^{N}e^{i\vec{k}\cdot(\vec{r}_{l}-\vec{r}_{m}%
)}\times\left\vert \phi_{0}\left(  \vec{r}_{1},\cdots,\vec{r}_{N}\right)
\right\vert ^{2}\nonumber\\
&  =%
{\displaystyle\int}
d^{3N}\vec{r}\;\left\vert \psi_{\vec{k}}\left(  \vec{r}_{1},\cdots,\vec{r}%
_{N}\right)  \right\vert ^{2}=\left\langle \psi_{\vec{k}}\right.  \left\vert
\psi_{\vec{k}}\right\rangle .
\end{align}
Since
\begin{align}
&  \left\langle \psi_{\vec{k}}\right\vert H-E_{0}\left\vert \psi_{\vec{k}%
}\right\rangle \nonumber\\
&  =-\frac{1}{2m}%
{\displaystyle\int}
d^{3N}\vec{r}\;\sum_{j=1}^{N}\left(  \vec{\nabla}_{j}e^{-i\vec{k}\cdot\vec
{r}_{j}}\right)  ^{2}\times\left\vert \phi_{0}\left(  \vec{r}_{1},\cdots
,\vec{r}_{N}\right)  \right\vert ^{2}=N\frac{k^{2}}{2m},
\end{align}
we get%
\begin{equation}
\frac{\left\langle \psi_{\vec{k}}\right\vert H-E_{0}\left\vert \psi_{\vec{k}%
}\right\rangle }{\left\langle \psi_{\vec{k}}\right.  \left\vert \psi_{\vec{k}%
}\right\rangle }=\frac{k^{2}}{2mS(k)}.
\end{equation}
By the variational principle, the energy $E(k)$ for the lowest excitation with
momentum $\vec{k}$ satisfies the upper bound%
\begin{equation}
E(k)\leq\frac{k^{2}}{2mS(k)}. \label{upperbound}%
\end{equation}
This upper bound applies to phonons at small $k$ as well rotons at larger $k$.
\ For a relatively dense system such as $^{4}$He a maximum in $S(k)$ occurs
near $k\approx2\pi/d,$ where $d$ is the average spacing between bosons.
\ Therefore a minimum in the upper bound\ (\ref{upperbound}) occurs at
$k\approx2\pi/d$, and this may also be reflected in $E(k)$. \ On the other
hand for very dilute Bose systems local spatial correlations are weaker and
become significant only at distance scales comparable to the scattering
length. \ In this case the maximum in $S(k)$ has been shown to occur at
$k\approx8/(\pi a_{\text{scatt}})$ \cite{Steinhauer:2003}.

The simple estimate $k_{R}\approx2\pi/d$ works rather well for superfluid
$^{4}$He. \ The particle density of superfluid $^{4}$He at $T=1.25\,$K and
$P=1.0\,$atm has been measured to be \cite{Keesom:1942}%

\begin{equation}
\frac{N}{V}\approx2.20\times10^{-2}\text{\AA }^{-3}.
\end{equation}
Making a rough estimate for the average spacing%
\begin{equation}
d\approx\left(  \frac{N}{V}\right)  ^{-1/3},
\end{equation}
we get the prediction%
\begin{equation}
k_{R}^{\text{theory}}\approx\frac{2\pi}{\left(  \frac{N}{V}\right)  ^{-1/3}%
}=1.76\text{\AA }^{-1}\text{,}%
\end{equation}
which is close to the direct experimental measurement of the roton minimum at
$T=1.26\,$K and $P=1.00$ atm \cite{Dietrich:1972PRA}$,$%
\begin{equation}
k_{R}^{\text{exp}}=1.902\text{\AA }^{-1}.
\end{equation}
\ 

We return now to our fermionic spin-1/2 system with $N$ up spins and $N$ down
spins at unitarity. \ Let $d$ be the average distance between neighboring
pairs of particles. \ We use the same approximation,%
\begin{equation}
d\approx\left(  \frac{N}{V}\right)  ^{-1/3}=N^{-1/3}L,
\end{equation}
and find that%
\begin{equation}
k_{R}\approx\frac{2\pi}{N^{-1/3}L}=\left(  \frac{4\pi}{3}\right)  ^{1/3}%
k_{F}\approx1.61k_{F}.
\end{equation}
Unfortunately this does not describe the minimum of $(E_{1}-E_{0})/E_{F}$ at
$k\approx0.8k_{F}$. \ In fact the minimum in $(E_{1}-E_{0})/E_{F}$ is very
close to one-half of $k_{R}$. \ This would appear to rule out the
interpretation as two rotons.

It is likely that our systems of $6-26$ particles have too few particles to
get an accurate reading for the excitation spectrum. \ Indeed there are not
enough particles to probe the small $k/k_{F}$ behavior where the lowest
excitations are described by phonons. \ Perhaps this low-energy excitation
changes character and goes up in energy as we include more particles and long
wavelength phonons\ emerge. \ This is possible. \ However we should still
answer the question of what is happening in the $6-26$ particle system and how
it is able to produce an excitation this low in energy.

\section{Two-particle correlations}

In order to study the properties of the ground state and the unknown
excitation in further detail we compute two-particle correlations. \ For the
same $N,N$ system with total momentum $\vec{P}=\vec{0}$ and total spin $S=0$,
we define the opposite-spin two-particle correlation function,%
\begin{equation}
\rho_{\downarrow\uparrow}(\vec{r},t_{1},t_{2})=\rho_{\uparrow\downarrow}%
(\vec{r},t_{1},t_{2})=\frac{\left\langle \Psi(t_{1})\right\vert a_{\uparrow
}^{\dagger}(\vec{r})a_{\uparrow}(\vec{r})a_{\downarrow}^{\dagger}(\vec
{0})a_{\downarrow}(\vec{0})\left\vert \Psi(t_{2})\right\rangle }{\left\langle
\Psi(t_{1})\right.  \left\vert \Psi(t_{2})\right\rangle },
\end{equation}
and the same-spin two-particle correlation function,%
\begin{equation}
\rho_{\uparrow\uparrow}(\vec{r},t_{1},t_{2})=\rho_{\downarrow\downarrow}%
(\vec{r},t_{1},t_{2})=\frac{\left\langle \Psi(t_{1})\right\vert :a_{\downarrow
}^{\dagger}(\vec{r})a_{\downarrow}(\vec{r})a_{\downarrow}^{\dagger}(\vec
{0})a_{\downarrow}(\vec{0}):\left\vert \Psi(t_{2})\right\rangle }{\left\langle
\Psi(t_{1})\right.  \left\vert \Psi(t_{2})\right\rangle }.
\end{equation}
In the auxiliary field transfer matrix formalism we compute these correlation
functions using%
\begin{equation}
\rho_{\uparrow\downarrow}(\vec{n}_{s},t_{1},t_{2})=\frac{%
{\displaystyle\int}
Ds\;\left\langle \Psi(t_{1}),s\right\vert a_{\uparrow}^{\dagger}(\vec{n}%
_{s})a_{\uparrow}(\vec{n}_{s})a_{\downarrow}^{\dagger}(\vec{0})a_{\downarrow
}(\vec{0})\left\vert \Psi(t_{2}),s\right\rangle \exp\left\{  -\frac{1}{2}%
\sum_{\vec{n}}\left[  s(\vec{n})\right]  ^{2}\right\}  }{%
{\displaystyle\int}
Ds\;\left\langle \Psi(t_{1}),s\right.  \left\vert \Psi(t_{2}),s\right\rangle
\exp\left\{  -\frac{1}{2}\sum_{\vec{n}}\left[  s(\vec{n})\right]
^{2}\right\}  }, \label{rho_up_down}%
\end{equation}
and%
\begin{equation}
\rho_{\downarrow\downarrow}(\vec{n}_{s},t_{1},t_{2})=\frac{%
{\displaystyle\int}
Ds\;\left\langle \Psi(t_{1}),s\right\vert :a_{\downarrow}^{\dagger}(\vec
{n}_{s})a_{\downarrow}(\vec{n}_{s})a_{\downarrow}^{\dagger}(\vec
{0})a_{\downarrow}(\vec{0}):\left\vert \Psi(t_{2}),s\right\rangle \exp\left\{
-\frac{1}{2}\sum_{\vec{n}}\left[  s(\vec{n})\right]  ^{2}\right\}  }{%
{\displaystyle\int}
Ds\;\left\langle \Psi(t_{1}),s\right.  \left\vert \Psi(t_{2}),s\right\rangle
\exp\left\{  -\frac{1}{2}\sum_{\vec{n}}\left[  s(\vec{n})\right]
^{2}\right\}  }. \label{rho_down_down}%
\end{equation}

Rather than calculating matrix elements of the two-particle operators in
(\ref{rho_up_down}) and (\ref{rho_down_down}) directly, we define%
\begin{equation}
M(\epsilon_{\uparrow},\epsilon_{\downarrow},\delta_{\downarrow})=:\exp\left[
\epsilon_{\uparrow}a_{\uparrow}^{\dag}(\vec{n}_{s})a_{\uparrow}(\vec{n}%
_{s})+\epsilon_{\downarrow}a_{\downarrow}^{\dag}(\vec{n}_{s})a_{\downarrow
}(\vec{n}_{s})+\delta_{\downarrow}a_{\downarrow}^{\dag}(\vec{0})a_{\downarrow
}(\vec{0})\right]  :.
\end{equation}
We extract the operators needed for the two-particle correlations by taking
derivatives,%
\begin{equation}
a_{\uparrow}^{\dagger}(\vec{n}_{s})a_{\uparrow}(\vec{n}_{s})a_{\downarrow
}^{\dagger}(\vec{0})a_{\downarrow}(\vec{0})=\lim_{\epsilon_{\uparrow}%
,\epsilon_{\downarrow},\delta_{\downarrow}\rightarrow0}\frac{\partial
}{\partial\epsilon_{\uparrow}}\frac{\partial}{\partial\delta_{\downarrow}%
}M(\epsilon_{\uparrow},\epsilon_{\downarrow},\delta_{\downarrow}),
\end{equation}%
\begin{equation}
:a_{\downarrow}^{\dagger}(\vec{n}_{s})a_{\downarrow}(\vec{n}_{s}%
)a_{\downarrow}^{\dagger}(\vec{0})a_{\downarrow}(\vec{0}):=\lim_{\epsilon
_{\uparrow},\epsilon_{\downarrow},\delta_{\downarrow}\rightarrow0}%
\frac{\partial}{\partial\epsilon_{\downarrow}}\frac{\partial}{\partial
\delta_{\downarrow}}M(\epsilon_{\uparrow},\epsilon_{\downarrow},\delta
_{\downarrow}).
\end{equation}
This construction is useful because $M(\epsilon_{\uparrow},\epsilon
_{\downarrow},\delta_{\downarrow})$ itself looks like a transfer matrix with
only single-nucleon operators.

Let us define the new single-particle matrix elements with one insertion of
$M^{X}(\epsilon_{\uparrow},\epsilon_{\downarrow},\delta_{\downarrow})$,%
\begin{align}
&  M_{ij}(s,t_{1},t_{2},\epsilon_{\uparrow},\epsilon_{\downarrow}%
,\delta_{\downarrow})\nonumber\\
&  =\left\langle p_{i}^{X}\right\vert M_{n_{t_{1}}+n_{t_{2}}-1}^{X}%
(s)\times\cdots\times M_{n_{t_{2}}}^{X}(s)M^{X}(\epsilon_{\uparrow}%
,\epsilon_{\downarrow},\delta_{\downarrow})M_{n_{t_{2}}-1}^{X}(s)\times
\cdots\times M_{0}^{X}(s)\left\vert p_{j}^{X}\right\rangle ,
\end{align}
We let%
\begin{equation}
\rho(t_{1},t_{2},\epsilon_{\uparrow},\epsilon_{\downarrow},\delta_{\downarrow
})=\frac{%
{\displaystyle\int}
Ds\;\exp\left\{  -\frac{1}{2}\sum_{\vec{n}}\left[  s(\vec{n})\right]
^{2}\right\}  \det M(s,t_{1},t_{2},\epsilon_{\uparrow},\epsilon_{\downarrow
},\delta_{\downarrow})}{%
{\displaystyle\int}
Ds\;\exp\left\{  -\frac{1}{2}\sum_{\vec{n}}\left[  s(\vec{n})\right]
^{2}\right\}  \det M(s,t_{1}+t_{2})}.
\end{equation}
Then%
\begin{equation}
\rho_{\uparrow\downarrow}(\vec{n}_{s},t_{1},t_{2})=\lim_{\epsilon_{\uparrow
},\epsilon_{\downarrow},\delta_{\downarrow}\rightarrow0}\frac{\partial
}{\partial\epsilon_{\uparrow}}\frac{\partial}{\partial\delta_{\downarrow}}%
\rho(t_{1},t_{2},\epsilon_{\uparrow},\epsilon_{\downarrow},\delta_{\downarrow
}).
\end{equation}%
\begin{equation}
\rho_{\downarrow\downarrow}(\vec{n}_{s},t_{1},t_{2})=\lim_{\epsilon_{\uparrow
},\epsilon_{\downarrow},\delta_{\downarrow}\rightarrow0}\frac{\partial
}{\partial\epsilon_{\downarrow}}\frac{\partial}{\partial\delta_{\downarrow}%
}\rho(t_{1},t_{2},\epsilon_{\uparrow},\epsilon_{\downarrow},\delta
_{\downarrow}),
\end{equation}

\section{Results for $\rho_{\uparrow\downarrow}$ and $\rho_{\downarrow
\downarrow}$}

We compute $\rho_{\uparrow\downarrow}(\vec{n}_{s},t/2,t/2)$ and $\rho
_{\downarrow\downarrow}(\vec{n}_{s},t/2,t/2)$ for $N=5$ and $L=4$ using the
same initial state $\left\vert \Psi_{0}^{\text{free}}\right\rangle $ used
previously to calculate $G_{\psi^{2}}^{\text{bare}}(\vec{n}_{s},t/2,t/2)$.
\ We recall that our initial state for $N=5$ corresponds with single particle
momenta $\left\langle 0,0,0\right\rangle $, $\left\langle \frac{2\pi}%
{L},0,0\right\rangle $, $\left\langle -\frac{2\pi}{L},0,0\right\rangle $,
$\left\langle 0,\frac{2\pi}{L},0\right\rangle $, and $\left\langle
0,-\frac{2\pi}{L},0\right\rangle $. \ For each lattice calculation we have
performed several consistency checks for $\rho_{\uparrow\downarrow}(\vec
{n}_{s},t/2,t/2)$ and $\rho_{\downarrow\downarrow}(\vec{n}_{s},t/2,t/2)$.
\ Summing $\rho_{\uparrow\downarrow}(\vec{n}_{s},t/2,t/2)$ over $\vec{n}_{s}$
and multiplying by $L^{3}$ counts the number of ways to select two particles
with opposite spins,%
\begin{equation}
L^{3}%
{\displaystyle\sum\limits_{\vec{n}_{s}}}
\rho_{\uparrow\downarrow}(\vec{n}_{s},t_{1},t_{2})=N^{2}.
\label{opposite_spin_check}%
\end{equation}
The same procedure for $\rho_{\uparrow\downarrow}(\vec{n}_{s},t/2,t/2)$ counts
the number of ways to select two down-spin particles without replacement,%
\begin{equation}
L^{3}%
{\displaystyle\sum\limits_{\vec{n}_{s}}}
\rho_{\downarrow\downarrow}(\vec{n}_{s},t/2,t/2)=N(N-1).
\label{same_spin_check}%
\end{equation}
One last consistency check uses the fact that%
\begin{equation}
a_{\uparrow}^{\dagger}(\vec{0})a_{\uparrow}(\vec{0})a_{\downarrow}^{\dagger
}(\vec{0})a_{\downarrow}(\vec{0})=a_{\downarrow}^{\dagger}(\vec{0}%
)a_{\uparrow}^{\dagger}(\vec{0})a_{\uparrow}(\vec{0})a_{\downarrow}(\vec{0}).
\label{origin_check}%
\end{equation}
From this we deduce that
\begin{equation}
\rho_{\uparrow\downarrow}(\vec{0},t/2,t/2)=G_{\psi^{2}}^{\text{bare}}(\vec
{0},t/2,t/2).
\end{equation}
The results of the three consistency checks are shown in Table 6. \ We see
that all ratios are in agreement with the required value of $1$.%
\[%
\genfrac{}{}{0pt}{}{%
\begin{tabular}
[c]{|c|c|c|c|}\hline
$E_{F}t$ & $\frac{L^{3}}{N^{2}}\sum\limits_{\vec{n}_{s}}\rho_{\uparrow
\downarrow}(\vec{n}_{s},t/2,t/2)$ & $\frac{L^{3}}{N(N-1)}\sum\limits_{\vec
{n}_{s}}\rho_{\downarrow\downarrow}(\vec{n}_{s},t/2,t/2)$ & $\frac
{\rho_{\uparrow\downarrow}(\vec{0},t/2,t/2)}{G_{\psi^{2}}^{\text{bare}}%
(\vec{0},t/2,t/2)}$\\\hline
$0$ & $1.00$ & $1.00$ & $1.00$\\\hline
$1.23$ & $1.00(1)$ & $1.00(1)$ & $1.00(2)$\\\hline
$2.46$ & $1.00(1)$ & $1.00(1)$ & $1.01(1)$\\\hline
$3.70$ & $1.00(1)$ & $1.00(1)$ & $1.01(1)$\\\hline
$4.93$ & $0.99(1)$ & $0.99(1)$ & $0.99(2)$\\\hline
$6.16$ & $1.01(1)$ & $1.01(1)$ & $1.02(1)$\\\hline
$7.39$ & $0.98(2)$ & $0.98(2)$ & $0.98(3)$\\\hline
\end{tabular}
\ \ }{\text{Table 6: \ Consistency checks for }\rho_{\uparrow\downarrow}\text{
and }\rho_{\downarrow\downarrow}.\text{ \ Each entry should equal }1\text{.}}%
\]

\begin{figure}
[ptb]
\begin{center}
\includegraphics[
height=2.6489in,
width=3.8346in
]%
{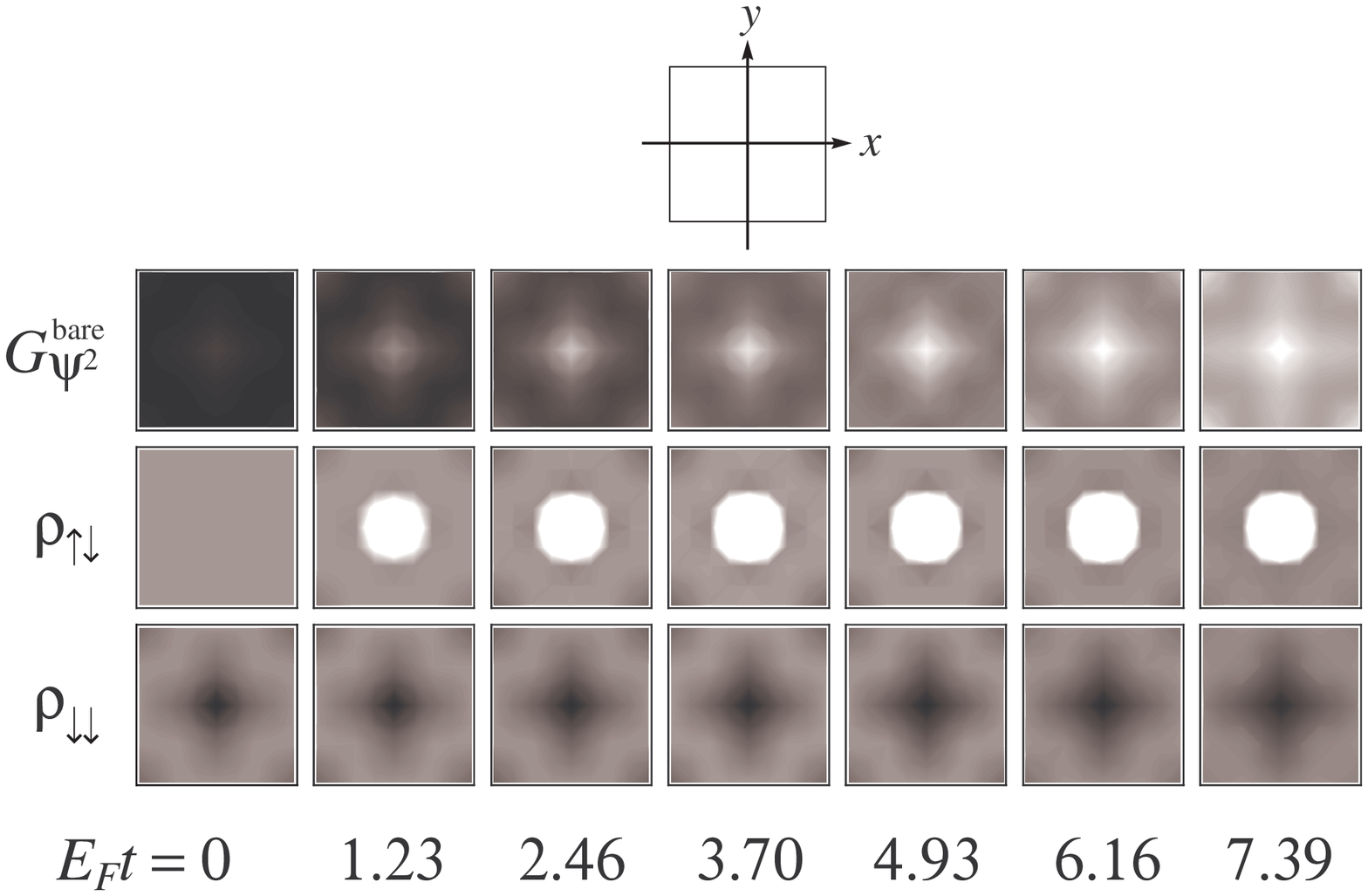}%
\caption{$G_{\psi^{2}}^{\text{bare}}(\vec{n}_{s},t/2,t/2)$, $\rho
_{\uparrow\downarrow}(\vec{n}_{s},t/2,t/2)$, $\rho_{\downarrow\downarrow}%
(\vec{n}_{s},t/2,t/2)$ with $\vec{n}_{s}$ in the $xy$-plane for $N=5$ and
$L=4$.}%
\label{l4_xy_labelled}%
\end{center}
\end{figure}

\begin{figure}
[ptbptb]
\begin{center}
\includegraphics[
height=2.6489in,
width=3.8346in
]%
{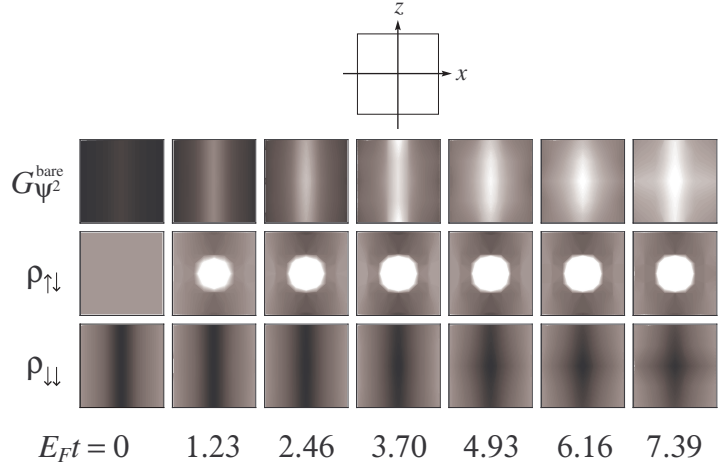}%
\caption{$G_{\psi^{2}}^{\text{bare}}(\vec{n}_{s},t/2,t/2)$, $\rho
_{\uparrow\downarrow}(\vec{n}_{s},t/2,t/2)$, $\rho_{\downarrow\downarrow}%
(\vec{n}_{s},t/2,t/2)$ with $\vec{n}_{s}$ in the $xz$-plane for $N=5$ and
$L=4$.}%
\label{l4_xz_labelled}%
\end{center}
\end{figure}

We show results for $G_{\psi^{2}}^{\text{bare}}(\vec{n}_{s},t/2,t/2)$,
$\rho_{\uparrow\downarrow}(\vec{n}_{s},t/2,t/2)$, and $\rho_{\downarrow
\downarrow}(\vec{n}_{s},t/2,t/2)$ with $\vec{n}_{s}$ in the $xy$-plane in FIG.
\ref{l4_xy_labelled}. \ The same results for $\vec{n}_{s}$ in the $xz$-plane
are shown FIG. \ref{l4_xz_labelled}. \ In our contour plots the maximum
brightness for $G_{\psi^{2}}^{\text{bare}}$ corresponds with $0.05$, while the
maximum brightness for $\rho_{\uparrow\downarrow}$ and $\rho_{\downarrow
\downarrow}$ corresponds with $0.01$.%
\ For $N=5$ the ground state of the free Fermi system is degenerate. \ We have
chosen the initial state $\left\vert \Psi_{0}^{\text{free}}\right\rangle $
with all momenta orthogonal to the $z$-direction. \ This explains the lack of
cubic $SO(3,\mathbb{Z})$ invariance in the $xz$-plane data, especially for
smaller values of $E_{F}t$. \ At the largest value $E_{F}t=7.39$ the
$xz$-plane data shows signs of cubic rotational invariance and is beginning to
resemble the $xy$-plane data at $E_{F}t=7.39$.

We see that already for $E_{F}t\geq1.23$ a sharp maximum in $\rho
_{\uparrow\downarrow}$ appears at $\vec{n}_{s}=0$. \ This indicates
significant spatial correlation between up-spin and down-spin particles due to
the strong zero-range attractive force. \ Another indication of this spatial
correlation is that for larger values of $E_{F}t$ and $\vec{n}_{s}$ away from
the origin, the shape and strength of $\rho_{\uparrow\downarrow}$ and
$\rho_{\downarrow\downarrow}$ are nearly identical. \ Another visible pattern
is that regions of maximum strength for $G_{\psi^{2}}^{\text{bare}}$ correlate
with regions of minimum strength for $\rho_{\downarrow\downarrow}$.
\ Conversely the minima for $G_{\psi^{2}}^{\text{bare}}$ are maxima in
$\rho_{\downarrow\downarrow}$. \ This is most likely the direct result of
Pauli blocking.

\begin{figure}
[ptb]
\begin{center}
\includegraphics[
height=1.7919in,
width=3.8553in
]%
{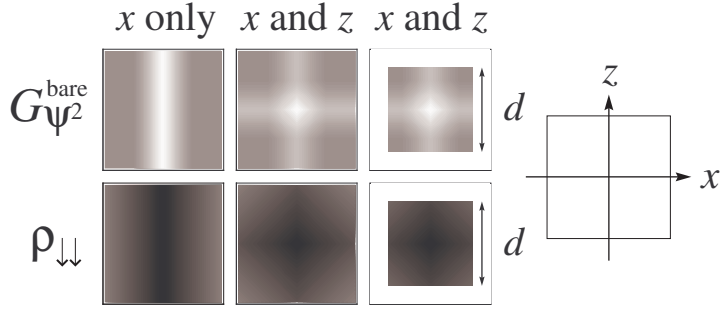}%
\caption{The first column shows $G_{\psi^{2}}^{\text{bare}}$ and
$\rho_{\downarrow\downarrow}$ for $\left\vert \Psi_{x}\right\rangle $. \ The
second column shows $G_{\psi^{2}}^{\text{bare}}$ and $\rho_{\downarrow
\downarrow}$ averaged for $\left\vert \Psi_{x}\right\rangle $ and $\left\vert
\Psi_{z}\right\rangle $. \ The third column shows the same data as the second
column but cropped to a region of size $d\times d$.}%
\label{1d_labelled}%
\end{center}
\end{figure}

The crosslike pattern that emerges in $G_{\psi^{2}}^{\text{bare}}$ and
$\rho_{\downarrow\downarrow}$ is the same $2\pi/L$ cosine wave excitation we
have been investigating. \ We note that the same pattern can be reproduced
rather simply with two up-spin and two down-spin particles. $\ $Let
$\left\vert \Psi_{x}\right\rangle $ be the normalized four particle state,%
\begin{equation}
\left\vert \Psi_{x}\right\rangle =\frac{1}{L^{6}\sqrt{2}}\left[  \tilde
{a}_{\uparrow}^{\dagger}(-\vec{p}_{x})\tilde{a}_{\downarrow}^{\dagger}(\vec
{p}_{x})+\tilde{a}_{\uparrow}^{\dagger}(\vec{p}_{x})\tilde{a}_{\downarrow
}^{\dagger}(-\vec{p}_{x})\right]  \tilde{a}_{\uparrow}^{\dagger}(\vec
{0})\tilde{a}_{\downarrow}^{\dagger}(\vec{0})\left\vert 0\right\rangle ,
\end{equation}%
\begin{equation}
\vec{p}_{x}=\left\langle \frac{2\pi}{L},0,0\right\rangle .
\end{equation}
Similarly let $\left\vert \Psi_{z}\right\rangle $ be%
\begin{equation}
\left\vert \Psi_{z}\right\rangle =\frac{1}{L^{6}\sqrt{2}}\left[  \tilde
{a}_{\uparrow}^{\dagger}(-\vec{p}_{z})\tilde{a}_{\downarrow}^{\dagger}(\vec
{p}_{z})+\tilde{a}_{\uparrow}^{\dagger}(\vec{p}_{z})\tilde{a}_{\downarrow
}^{\dagger}(-\vec{p}_{z})\right]  \tilde{a}_{\uparrow}^{\dagger}(\vec
{0})\tilde{a}_{\downarrow}^{\dagger}(\vec{0})\left\vert 0\right\rangle ,
\end{equation}%
\begin{equation}
\vec{p}_{z}=\left\langle 0,0,\frac{2\pi}{L}\right\rangle .
\end{equation}
We define $t$-independent expressions $G_{\psi^{2}}^{\text{bare}}(\vec{n}%
_{s},\Psi_{x,z})$ and $\rho_{\downarrow\downarrow}(\vec{n}_{s},\Psi_{x,z})$
for the states $\left\vert \Psi_{x}\right\rangle $ and $\left\vert \Psi
_{z}\right\rangle ,$%
\begin{align}
G_{\psi^{2}}^{\text{bare}}(\vec{n}_{s},\Psi_{x,z})  &  =\left\langle
\Psi_{x,z}\right\vert a_{\downarrow}^{\dagger}(\vec{n}_{s})a_{\uparrow}%
(\vec{n}_{s})a_{\uparrow}^{\dagger}(\vec{0})a_{\downarrow}(\vec{0})\left\vert
\Psi_{x,z}\right\rangle \\
\rho_{\downarrow\downarrow}(\vec{n}_{s},\Psi_{x,z})  &  =\left\langle
\Psi_{x,z}\right\vert :a_{\downarrow}^{\dagger}(\vec{n}_{s})a_{\downarrow
}(\vec{n}_{s})a_{\downarrow}^{\dagger}(\vec{0})a_{\downarrow}(\vec
{0}):\left\vert \Psi_{x,z}\right\rangle .
\end{align}
The first column in FIG. \ref{1d_labelled} shows $G_{\psi^{2}}^{\text{bare}}$
and $\rho_{\downarrow\downarrow}$ for $\left\vert \Psi_{x}\right\rangle $.
\ The second column shows $G_{\psi^{2}}^{\text{bare}}$ and $\rho
_{\downarrow\downarrow}$ averaged for $\left\vert \Psi_{x}\right\rangle $ and
$\left\vert \Psi_{z}\right\rangle $. \ The third column shows the same data as
the second column but cropped to a region of size $d\times d$, where $d\approx
N^{-1/3}L$ is the average spacing between particles. \ For these contour plots
the maximum brightness for $G_{\psi^{2}}^{\text{bare}}$ corresponds with
$0.00125$, while the maximum brightness for $\rho_{\downarrow\downarrow}$
corresponds with $0.001$. \ We note that the intensity scale for
$\rho_{\downarrow\downarrow}$ is a factor of ten lower than that in FIGS.
\ref{l4_xy_labelled} and \ref{l4_xz_labelled}. \ This is consistent with the
sum rule in (\ref{same_spin_check}) since we have reduced $N(N-1)$ from $20$
to $2$. \ The ratio of intensities between $G_{\psi^{2}}^{\text{bare}}$ and
$\rho_{\downarrow\downarrow}$ in FIGS. \ref{l4_xy_labelled} and
\ref{l4_xz_labelled} is a factor of four larger than that in FIG.
\ref{1d_labelled}. \ This implies that the actual excitation has stronger
spatial correlations between up and down spins than our simple plane-wave
states $\left\vert \Psi_{x}\right\rangle $ and $\left\vert \Psi_{z}%
\right\rangle $.%

We find that the $d\times d$ region in the third column of (\ref{1d_labelled})
describes the transient signal in $G_{\psi^{2}}^{\text{bare}}$ and
$\rho_{\downarrow\downarrow}$ for large $E_{F}t$. \ In fact this holds true
for all systems we have checked, $N=3,5,7,9,11,13$. \ This would suggest that
the excitation is some type of quasi-1D system of four particles wrapped
around the periodic lattice. \ The $d\times d$ window would suggest a
transverse width of approximately $d$ for the quasi-1D system. \ Outside the
$d\times d$ region it is more difficult to discern a universal pattern in the
data for all $N$.

\section{Summary and discussion}

We have presented lattice results for spin-1/2 fermions at unitarity, where
the effective range of the interaction is zero and the scattering length is
infinite. \ We measured the spatial coherence of difermion pairs for a system
of particles with equal numbers of up and down spins in a periodic cube.
\ Using a transfer matrix projection method with auxiliary fields, we analyzed
both ground state properties and transient behavior due to low-energy
excitations. \ We first measured $\Gamma,$ the probability that a given
lattice site has both an up-spin and down-spin particle. \ From this we were
able to deduce that $\xi_{1}=1.0(1),$ where $\xi_{1}$ is defined by%
\begin{equation}
\frac{E_{0}}{N+N}=\frac{3}{5}\frac{k_{F}^{2}}{2m}\left[  \xi-\xi_{1}k_{F}%
^{-1}a_{\text{scatt}}^{-1}+O(k_{F}^{-2}a_{\text{scatt}}^{-2})\right]  .
\end{equation}
We then measured the pair correlation function at nonzero spatial separation
and found clear evidence of off-diagonal long-range order. \ This suggests
that the ground state is superfluid with s-wave pairing.

We also found evidence for a low-energy excitation with energy $E_{1}$ which
ranges between $0.05E_{F}$ and $0.2E_{F}$ for $N=3,5,7,9,11,13$. \ We note
that recent radio frequency measurements of the excitation spectra in $^{6}$Li
at unitarity has found an energy gap in the unpolarized system at roughly
$0.2E_{F}$ \cite{Chin:2004}. \ The low value for the excitation energy would
tend to drive down the superfluid critical temperature and create a pseudogap
phase where single fermionic quasiparticles are gapped at $\Delta$ while the
new excitation is gapped at a lower energy scale. \ An estimate of the effect
on the critical temperature requires a better understanding of the energy gap
in large systems, the density of states, and possible interactions. \ It is
possible though that this excitation could explain the consistently low
critical temperature measured in recent lattice simulations which have
specifically looked for off-diagonal long-range order
\cite{Wingate:2005xy,Lee:2005it,Burovski:2006a,Burovski:2006b}. \ 

The low-energy excitation corresponding with $E_{1}$ in systems
$N=3,5,7,9,11,13$ appears to be inconsistent with an interpretation as a pair
of weakly interacting phonons, fermionic quasiparticles, or rotons. \ By
examining the pair correlation function as well as two-particle density
correlations, we find the excitation has the characteristics of a quasi-1D
chain consisting of two up-spin and two down-spin particles aligned along one
of the lattice axes. \ We caution that this excitation could be an artifact
due to the periodic boundary with a relatively small number of particles.
\ Confirmation in lattice systems with more particles and different geometries
will be needed. \ The exact mechanism which produces the quasi-1D chain is
beyond the scope of this study. \ However we can present here at least one
plausible explanation.

\begin{figure}
[ptb]
\begin{center}
\includegraphics[
height=1.5515in,
width=3.0856in
]%
{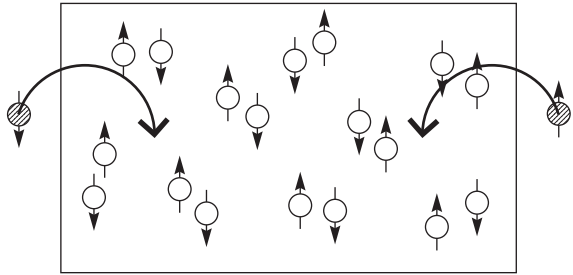}%
\caption{Starting with the ground state, we insert one extra up-spin and one
extra down-spin.}%
\label{insert_two}%
\end{center}
\end{figure}

Consider the ground state of the unitary limit system with $N-1$ up spins and
$N-1$ down spins. \ Let $d$ be the average separation between particles. \ We
now introduce one extra up-spin and one extra down-spin separated by a
distance $x>d$ as shown in FIG. \ref{insert_two}. \ Next we evolve the state
forward in Euclidean time using the projection operator $e^{-Ht}$. \ If
$x\gg\sqrt{t/m}\sim d$ then the projection time is not sufficient to find the
true ground state of the $N,N$ system. \ Instead we have two essentially
non-interacting quasiparticles, and the expectation value of the energy after
Euclidean time $t$ will be roughly $2\Delta$ above the ground state energy,
where $\Delta$ is the even-odd pairing gap.%

\begin{figure}
[ptb]
\begin{center}
\includegraphics[
height=1.5515in,
width=2.5278in
]%
{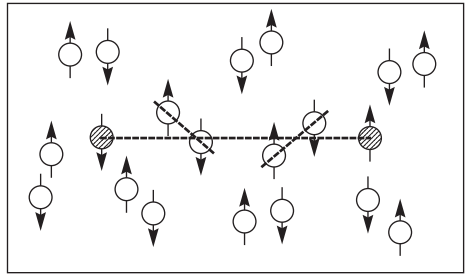}%
\caption{The pairs located near a line segment connecting the extra particles
favor an orientation with the up spins to the left and down spins to the
right.}%
\label{oriented}%
\end{center}
\end{figure}

If however $x\gtrsim\sqrt{t/m}\sim d$, then pairs nearby the extra particles
may rearrange themselves slightly to lower the total energy as shown in FIG.
\ref{oriented}. \ Pairs located near a line segment connecting the extra
particles favor an orientation with up spins to the left and down spins to the
right. \ The tilt angle with respect to the line segment may prefer to
alternate from one pair to the next. \ The resulting state could be described
as a pair of quasiparticles interacting via a loose chain of pairs connecting
them. \ The transverse width of this chain would be roughly equal to $d$.%

\begin{figure}
[ptb]
\begin{center}
\includegraphics[
height=1.0032in,
width=2.5002in
]%
{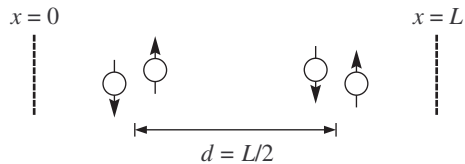}%
\caption{Two up spins and two down spins aligned along one of the lattice
axes.}%
\label{two_pairs_l}%
\end{center}
\end{figure}

Chain configurations without specified endpoints can also be constructed. \ In
FIG. \ref{two_pairs_l} we show a four-particle chain extending across the
periodic lattice. \ From the symmetry of the four-particle chain configuration
we expect the resonance energy to be minimized when the average spacing
between particles equals $L/2$.%
Using a simple estimate for the average separation, $d\approx L/N^{1/3}$, we
find that a minimum in energy should occur at $N\approx8$. \ This is
consistent with the $N$-dependence of $(E_{1}-E_{0})/E_{F}$ shown in FIG.
\ref{dispersion}. \ Further studies will be needed to see if longer quasi-1D
chains exist and if excitations can been found which do not wind around the
periodic lattice. \ If longer chains are possible then we might expect
resonances in a periodic cube for $d=L/j$ for each integer $j\geq2$. \ This
corresponds with $N\approx j^{3}$. \ However the alternation of tilt angle
between neighboring pairs will be frustrated for odd $j$

In our discussion we have tried to put an emphasis on universal observables in
the unitary limit. \ These are observables which agree in different physical
systems at unitarity, such as critical temperature, critical velocity, speed
of sound, low-energy excitation energies, etc. \ The reason for the emphasis
on universal observables is that different microscopic theories can agree on
all low-energy physical observables but disagree on wavefunctions and
off-shell Green's functions. \ In order to make contact with recent
observations of superfluidity with trapped ultracold atoms
\cite{Regal:2004,Kinast:2004,Zwierlein:2004,Bartenstein:2004,Chin:2004,Zwierlein:2005}%
, it seems important for both theory and experiment to identify truly
universal properties of the unitary limit. \ There has been some recent work
on the non-universality of fermion momentum occupation numbers with respect to
field redefinitions \cite{Furnstahl:2000we,Furnstahl:2001xq}. \ By analogy
similar issues have been raised concerning the universality of superfluid
condensate fractions for a low-energy effective theory without specification
of the fundamental microscopic physics.

\section{Acknowledgements}

The author is grateful to Gautam Rupak and Thomas Sch\"{a}fer for many
stimulating discussions. \ This work is supported in part by the DOE grant DE-FG02-04ER41335.

\bibliographystyle{apsrev}
\bibliography{NuclearMatter}

\end{document}